\documentclass[prb,twocolumn,showpacs,preprintnumbers,amsmath,amssymb,superscri
taddress]{revtex4}
\usepackage{graphicx}
\usepackage{dcolumn}
\usepackage{bm}

\begin{document}
\title{Doping dependence of spin excitations in the stripe phase of high-$T_c$
superconductors}
\author{G. Seibold}
\affiliation{Institut f\"ur Physik, BTU Cottbus, PBox 101344,
         03013 Cottbus, Germany}
\author{J. Lorenzana}
\affiliation{SMC-INFM,ISC-CNR, Dipartimento di Fisica,
Universit\`a di Roma La Sapienza, P. Aldo Moro 2, 00185 Roma, Italy}
\date{\today}

\begin{abstract}
Based on the time-dependent Gutzwiller approximation for the
extended Hubbard model we calculate the energy and momentum
dependence of spin excitations for striped ground states. 
Our starting point correctly reproduces the observed doping dependence
of the incommensurability in La-based cuprates and the
dispersion  of magnetic modes in 
the insulating parent compound. This allows us to make quantitative
predictions for the doping evolution of the dispersion of magnetic 
modes in the stripe phase including the energy $\omega_0$ 
and intensity of the resonance peak as well as the velocity $c$ of the
spin-wave like Goldstone mode.
In the underdoped regime $n_h<1/8$ we find a weak linear
dependence of $\omega_0$ on doping whereas the resonance
energy significantly shifts to higher values when the charge
concentration in the stripes starts to deviate from half-filling
for $n_h>1/8$. The velocity $c$ is non-monotonous with a minimum at
$1/8$ in coincidence with a well known anomaly in $T_c$.
Our calculations are in good agreement with available experimental data.
We also compare our results with analogous computations
based on linear spin-wave theory.
\end{abstract}
\maketitle
\section{Introduction}
The development of a theory for high-$T_c$
superconductivity presupposes a detailed understanding of the
normal state collective excitations in the charge and magnetic
channels. This paper is devoted to the latter channel
where we explore the spectrum of spin fluctuations based
on a stripe scenario extending the results presented 
in a previous short communication.\cite{sei05}

The large variety of cuprate materials has
often narrowed the view on the properties of individual compounds
rather than the common features between them. 
In YBa$_2$Cu$_3$O$_y$ (YBCO) compounds inelastic neutron scattering 
(INS) experiments have usually  
emphasized the  ``resonance mode'' \cite{ros91,pbou96,dai99,bou02,pai04},
an inelastic feature that sharpens as a delta-like 
excitation at frequency $\omega_0$ below $T_c$ and is located 
around the commensurate antiferromagnetic wave vector
${\bf Q}_{AF}=(\pi,\pi)$ (we set the lattice constant $a\equiv 1$ and
restore it when convenient for clarity).
On the other hand in lanthanum based cuprates (LCO) 
the emphasis has been put on
incommensurate features, either elastic or low-energy inelastic
as for example Goldstone like modes emerging from incommensurate 
wave vectors.\cite{tra95,tra96,tra97,yam98,tra04,fuj04,chr04}

In LCO the incommensurate low-energy magnetic scattering occurs at
wave vectors  ${\bf Q}_s=(\pi \pm 2\pi \epsilon,\pi )  $ and   
${\bf Q}_s=(\pi,\pi \pm 2\pi \epsilon ) $ and  $\epsilon$
depends linearly on doping $\epsilon=n_h$ up to $n_h \approx
1/8$. \cite{yam98} 
Here $n_h$ is the number of
added holes per planar Cu with respect to the parent insulating compound. 
This behavior is compatible with stripe like modulations of charge and  spin 
having a linear concentration of added holes $\nu=1/2$ (so-called
half-filled stripes), that is $\sim$ 1 hole per every 2nd elementary unit cell 
along the stripe and with charge (spin) periodicity $d$ ($2d$)
perpendicular to the stripe so that $\epsilon=1/(2d)$ and
$n_h=\nu/d$. 

Regarding superconductivity Wakimoto and collaborators\cite{wak04}
have recently found a direct
relation between $T_c$ and the intensity of low energy scattering in the
overdoped regime that together with other studies\cite{wak99} establishes 
a direct relation between superconductivity and incommensurate scattering
parallel to the Cu-O bond across the full phase diagram of LCO
compounds and makes more stringent the understanding of the 
magnetic excitations.

Despite the different emphasis mentioned above, 
some investigations in the last years   
started to focus on common features suggesting a universal 
phenomenology in the spin dynamics between different classes of 
high-T$_c$ cuprates. 
These similarities comprise the low-energy incommensurate
scattering in YBCO\cite{dai01} which has a similar dependence 
on doping as in LCO, namely a linear increase at small doping followed by a
saturation at larger concentrations. Even static incommensurate order has been 
reported\cite{moo02} in strongly underdoped YBCO.
Moreover, in detwinned samples of YBCO the incommensurate 
scattering has been shown to be anisotropic suggesting underlying stripe
correlations.\cite{moo00nat}  The presence of a ``rigid stripe array'' 
has been questioned in the more recent Ref.~\onlinecite{hin04}
which also reports an anisotropic response although in a more subtle way. 
We will come back to this point in Sec.~\ref{result}.

Finally, recent INS measurements show a remarkable similarity of the 
magnetic response of 
 La$_{2-x}$Ba$_{x}$CuO$_4$ (LBCO)\cite{tra04} and YBCO\cite{hay04}  
over a broad range of momenta and energy. 
Both experiments have revealed low energy incommensurate
excitations which, 
with increasing energy, continuously disperse
towards the resonance mode at ${\bf Q}_{AF}$.\cite{tra04,hay04}
An INS study on optimally doped La$_{2-x}$Sr$_{x}$CuO$_4$ (LSCO) 
\cite{chr04} has also resolved the dispersion of the 
spin-wave modes emerging from the incommensurate positions 
which closely resemble corresponding data of YBa$_2$Cu$_3$O$_{6.85}$.\cite{ron00}

Above $\omega_0$ the magnetic excitations start to form a ring or square 
shaped pattern around ${\bf Q}_{AF}$ which increases in size with 
increasing energy. In twinned samples low energy acoustic modes together 
with the high-energy
excitations thus display an 'X'-like dispersion. 
These features are now well established in both LCO 
(eventually codoped with Nd or Ba) \cite{tra04,chr04} and YBCO
\cite{ara99,pai04,hay04,hin04,stock04,rez04} while in Bi-
\cite{fon99} and Tl-materials \cite{hhe02} only the resonance peak
has been resolved yet.

The two most prominent explanations for the features in the
magnetic scattering are (a) a scenario based on
dispersing two-particle bound states induced by the antiferromagnetic (AFM) 
correlations in a d-wave superconducting system\cite{yin00,bri99,nor00,onu02,iik05,ito04,schny04,ere05a,erem205} and
(b) theories which rely on the presence of stripe correlations in
the ground state of the system.\cite{sei05,voj04,uhr04,uhrmag,uhr05,moe04,kru03,car04}
Approaches belonging to (a) are rather popular in describing spin excitations
in the superconducting state of YBCO materials. The ground state is
assumed to be an homogeneous metal from the start. 
\cite{yin00,bri99,nor00,onu02,iik05,ito04,schny04,ere05a}  Superconductivity
on top of this induces a spin-gap on the dynamical susceptibility and 
an RPA resummation of the AFM interaction creates exciton
peaks in the imaginary part of the magnetic susceptibility.
Due to the special frequency and momentum structure of the non-interacting
spin response functions in a d-wave superconducting system, 
the strongest weight is acquired by the 'resonance peak' at the 
AFM wave-vector and frequency $\omega_{0}$ close to the spin-gap energy. 
Above and below $\omega_{0}$ also an incommensurate response can be
obtained the structure of which strongly depends on the underlying
tight-binding dispersion. 
This 'excitonic' picture finds support in
the fact that in optimally doped YBCO  the resonance appears
below $T_c$ and offers  a natural explanation for the relation
$\omega_0 \sim  kT_c$ as observed in YBCO  
(see e.g. Ref.~\onlinecite{bou02}). However in underdoped 
YBCO the resonance is observed above $T_c$\cite{dai99} and  
upon applying this approach
to a quantitative description of magnetic excitations in LSCO 
it has been shown to be incompatible with
the low energy response.\cite{iik05,ito04} 
Moreover, in order to properly account for the
doping dependence of the low energy incommensurability in YBCO this 
approach has to rely on peculiar nesting properties of the
Fermi surface over a large doping range.\cite{notesherman}

The incommensurate correlations observed on several compounds have
encouraged theories based on the presence of stripe correlations 
in the ground state of the system.\cite{kane00,kane02,kru03,car04,
sei05,voj04,uhr04,uhr05,moe04} 
 The starting point is a broken symmetry state where
at least $C_4$ lattice symmetry is broken, translation
symmetry is usually broken (at least in explicit computations) and the
breaking of SU(2) spin rotational symmetry depends on the approach.  
If SU(2) is conserved as in Refs. \onlinecite{voj04,uhr04,uhr05} the 
system has a spin 
gap in the spectrum whereas if it is
broken,\cite{kane00,kane02,kru03,car04,sei05,moe04} as in the present
work, one finds a Goldstone magnetic mode emerging from the
incommensurate wave vector. A spin gap is observed in YBCO compounds
whereas a Goldstone like mode  is observed in  LCO. The response of
both systems is very similar above the gap.\cite{chr04} We will come back
to this point in Sec.~\ref{conclusion}.

In our SU(2) broken approach, at small doping, partially filled
stripes are stable\cite{whi98prl80,whi98prl81}
explaining the linear dependence of the incommensurability on 
doping for $n_h < 1/8$.\cite{fle00,fle01b,lor02b,sei04a} 
Our previous computations in the Gutzwiller approximation 
(GA)\cite{lor02b,sei04a} 
have shown that due to their transverse extension, stripes
with $d<4$ are energetically unfavorable so that for $n_h > 1/8$ 
the incommensurability remains at $\epsilon \approx 1/8$ in accord
with experiment and the stripe filling 
starts to increase beyond the value $\nu=1/2$. This picture is also in
agreement with dynamical mean-field theory 
computations.\cite{fle00,fle01b}

Within this scenario the resonance feature appears as a saddle-point 
(or a local maximum) 
in the dispersion relation of magnetic excitations as quantitatively
evaluated for LBCO in our previous work.\cite{sei05} 
The idea that the resonance may be seen as the lowest energy magnetic 
excitation at ${\bf Q}_{AF}$ in an incommensurate system was 
already proposed in Ref.~\onlinecite{bat01}.

Within the stripe scenario another line of thought  
has appeared based on simplified effective spin 
only models.  These studies  map the stripe structure to an array of 
coupled $n$-leg spin ladders with $n$ even. 
This allows to consider 
states where SU(2) is not broken and which exhibit a spin gap as in
YBCO.\cite{voj04,uhr04,uhr05,uhrmag} However, for some dopings $n$
is expected to be odd and the system should be gapless. 
We are not aware of evidence for this even/odd phenomenon.

In this paper we present detailed computations of the dynamical magnetic
structure factor on top of the static GA stripe textures using the
time dependent Gutzwiller approximation (TDGA).\cite{sei01,sei04b}
In addition we compare the TDGA results with linear spin-wave
theory (LSWT) computations.
We present the doping dependence of magnetic excitations 
extending our previous results\cite{sei05} for doping $n_h=1/8$. 
To be specific we choose parameters as appropriate
for LBCO or LSCO where our investigation
provides a quantitative prediction for the relation
between $\omega_0$, the spin-wave velocity, the 
dispersion of high energy modes, etc. as a function of doping $n_h$. 
As in our previous report\cite{sei05} parameters are fixed by fitting
the dispersion relation in the insulator and by requiring the correct
filling $\nu$ of the stripes. For the rest the computations in the doped phase
can be considered 
without free parameters and are in good agreement with data at
$n_h=1/8$\cite{tra04} on LBCO. We compare the new results with experimental 
data\cite{chr04} at $n_h=0.1$ and $n_h=0.16$. Encouraged by the
experimental finding of universality of spin excitations in
different cuprate materials we also attempt to make connection of our results
with experimental features observed in YBCO materials. In this case
however, due to the presence of the spin gap and the bilayer structure, 
our results will be more qualitative than quantitative.

This paper is organized as follows. In Sec.~\ref{modform} we
present the model and outline the formalism for the computation of
magnetic excitations. In Sec.~\ref{parameters} we show how the
appropriate parameter set can be obtained from experimental and
theoretical considerations. In addition we present  the
saddle-point solutions on which the subsequent RPA fluctuations are
computed. 
In Sec.~\ref{result} we
present results for the doping dependence of magnetic excitations
in the stripe phase and discuss these spectra with respect to the
present experimental situation. Finally we conclude our
considerations in Sec.~\ref{conclusion}.

\section{Model and Formalism}\label{modform}

Our investigations are based on the one-band Hubbard model with
hopping restricted to nearest ($t_{ij}=-t$) and next nearest ($t_{ij}=-t'$)
neighbors
\begin{equation}\label{HM}
H=\sum_{ij,\sigma}t_{ij}c_{i,\sigma}^{\dagger}c_{j,\sigma}
+ U\sum_{i}
n_{i,\uparrow}n_{i,\downarrow}.\nonumber
\end{equation}
Here $c_{i,\sigma}^{(\dagger)}$ destroys (creates) an electron
with spin $\sigma$ at site
$i$, and $n_{i,\sigma}=c_{i,\sigma}^{\dagger}c_{i,\sigma}$. $U$ is the
on-site Hubbard repulsion.

\subsection{Time-dependent Gutzwiller approximation}

\begin{figure}[bp]
\includegraphics[width=8.5cm,clip=true]{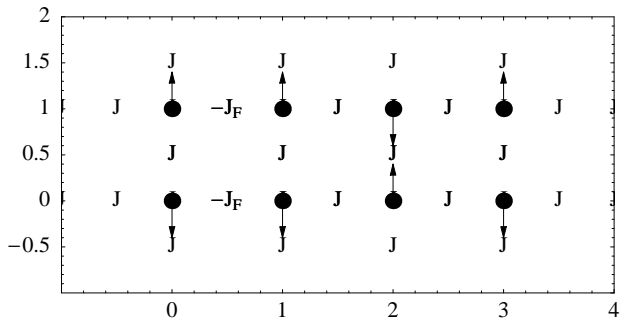}
\caption{Unit cell used for the $d=4$ stripe computation.  The length
 of the spin arrows corresponds to the LSWT computations.
(For the GA computations the spin magnitude is modulated as  shown in
Fig.~\ref{fig0}). We also show the interactions used in the LSWT
computation:  a uniform value of the AFM interaction $J$ across all the bonds 
except for the ferromagnetic bonds where the sign is opposite to
ensure stability and the modulus is $J_F=0.2J$. } \label{fig:ucell}
\end{figure}

As a starting point we treat the model Eq.~(\ref{HM}) 
within an unrestricted Gutzwiller
approximation (GA) in the spin-rotational invariant 
formulation.\cite{li89,fre92,bak98}  
Basically one constructs a Gutzwiller wave function $|\Psi\rangle$
by applying a projector to a Slater determinant $|SD\rangle$ which reduces
the double occupancy. The Slater determinant is allowed to have an
inhomogeneous charge and spin distribution describing generalized spin
and charge density waves determined variationally.\cite{geb90} 

The resulting energy functional $E^{GA}\equiv \langle \Psi | H| \Psi
\rangle $ treated in the Gutzwiller 
approximation reads as \cite{sei04b}
\begin{equation}\label{EGA}
E^{GA}= \sum_{i,j,\sigma,\sigma_1,\sigma_2}
t_{ij} z_{i,\sigma_1,\sigma}
z_{j,\sigma,\sigma_2} \langle c^\dagger_{i\sigma_1}c_{j\sigma_2}\rangle_0
+ U\sum_{i}D_{i},
\end{equation}
where the matrix ${\bf z}_i$ is given by
\begin{equation}\label{zmat}
{\bf z}_i=\left( \begin{array}{cc}
{z_i}^+\cos^2\frac{\Phi}{2}+{z_i}^-\sin^2\frac{\Phi}{2} &
\frac{S_i^-}{S_i^z}[{z_i}^+-{z_i}^-]\cos\Phi \\
\frac{S_i^+}{S_i^z}[{z_i}^+-{z_i}^-]\cos\Phi &
{z_i}^+\sin^2\frac{\Phi}{2}+{z_i}^-\cos^2\frac{\Phi}{2}
\end{array} \right),
\end{equation}
and
\begin{eqnarray*}
\tan^2\Phi&=&\frac{S_i^+S_i^-}{(S_i^z)^2},\\
{z_i}^\pm &=& \frac{\sqrt{1-n_{i}+D_i}\lambda_i^\pm
+\lambda_i^{\mp}\sqrt{D_i}}{\sqrt{\left(1-D_i-(\lambda_i^\pm)^2\right)
\left(n_{i}-D_i-(\lambda_i^\mp)^2\right)}},\\
(\lambda_i^\pm)^2 &=& n_i/2 - D_i \pm S_i^z\sqrt{1+\tan^2\Phi}.
\end{eqnarray*}

The above expressions are given in terms of the one-body density
matrix 
$\rho_{i\sigma,j\sigma'}\equiv \langle
c^\dagger_{i\sigma}c_{j\sigma'}\rangle_0$ 
in the  Slater determinant $|SD\rangle$.
For brevity we denote such averages with $  \langle ... \rangle_0$ and
introduce the notations:
$S_i^+\equiv \langle c^\dagger_{i\uparrow}c_{i\downarrow}\rangle_0$,
 $n_i \equiv \sum_\sigma\langle
 c^\dagger_{i\sigma}c_{i\sigma}\rangle_0$, etc.

In order to obtain generally inhomogeneous  solutions (stripes, etc.) 
one has to minimize Eq.~\ref{EGA} with respect to the double occupancy 
parameters $D_i=\langle \Psi | n_{i\uparrow}n_{i\downarrow}| \Psi
\rangle$ and with respect to  $\rho$  under
the constraint that $\rho$
corresponds to a Slater determinant.
For simplicity we restrict our saddle-point solutions to the limit
$S_i^\pm=0$, where the matrix ${\bf z}_i$ is
diagonal and the standard Gutzwiller energy functional
as derived by Gebhard\cite{geb90} or
Kotliar and Ruckenstein is recovered.\cite{kot86}
(The spin-rotational invariant formulation is necessary for the 
fluctuation computation).

We have previously shown that with realistic parameters 
this approach leads to vertical, metallic stripes in agreement 
with experimental data.\cite{lor02b,sei04a} 
Here we have repeated these calculations in large systems
typically consisting of $N\sim 100 \times 100$ lattice sites in order 
to obtain sufficient momentum resolution for the computation of
spin excitations. For stripes oriented along the $y$-direction this is 
most conveniently done by decomposing
the lattice into $N_{cells}$ unit cells of size $N_a=d\times 2$  with 
$N=N_{cells} N_a$. The unit cell for 
the $d=4$ solution is schematized  in Fig.~\ref{fig:ucell}. The 
locations of the unit cells define a 
Bravais ``superlattice''.\cite{brabais} The positions of the 
Bravais superlattice are given by 
${\bf R}\equiv n_1 {\bf a}_1 + n_2 {\bf a}_2 $ with $n_i$ integer.
For even $d$ the Bravais lattice is generated by the elementary
translations ${\bf a}_1=(d,1)$
and ${\bf a}_2 =(0,2)$ whereas for $d$ being odd we have ${\bf a}_1=(d,0)$ and ${\bf a}_2=(0,2)$.  
From the translation vectors
we can obtain the generators ${\bf b}_i$ of the reciprocal superlattice
defined by the relation  ${\bf a}_i.{\bf b}_j =2\pi\delta_{ij}$
and construct the associated magnetic Brillouin zone (MBZ). 
For odd $d$ the latter is simply a rectangle 
spanned by the reciprocal superlattice vectors ${\bf b}_1=(2\pi/d,0)$ and
${\bf b}_2=(0,\pi)$
whereas for even $d$ the shape is more complex due to the
non-orthogonality of the translation vectors.  

\begin{figure}[tbp]
\includegraphics[width=8.5cm,clip=true]{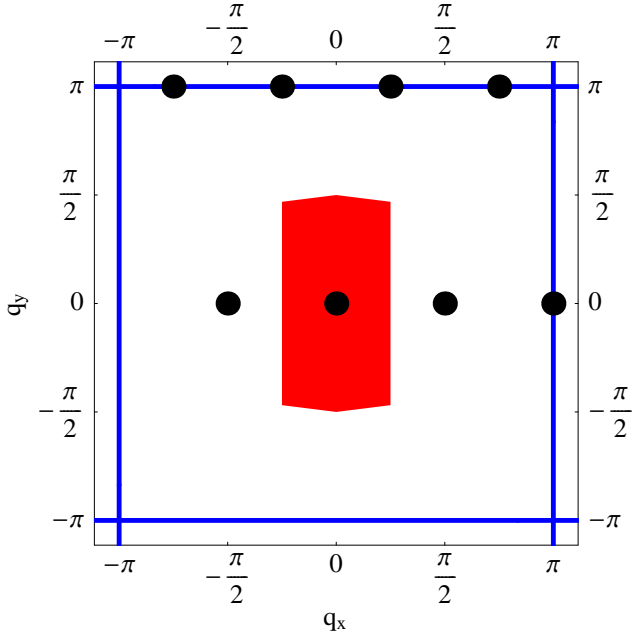}
\caption{(Color online) The central polygon is the first 
  magnetic Brillouin
  zone for $d=4$ stripes. Thick lines enclose the extended
  Brillouin zone. The dots are a set of magnetic
 reciprocal superlattice vectors that define a set of higher magnetic Brillouin
  zones with total volume equal to the EBZ volume. 
Magnetic modes and single particle states 
can be classified by the momentum in the central magnetic 
Brillouin zone and a band index. The  magnetic
 reciprocal superlattice vectors closest to $(\pm\pi,\pm\pi)$ 
are referred to as $Q^s$.} \label{fig:pbz}
\end{figure}

As an example we show in Fig.~\ref{fig:pbz} the first magnetic Brillouin
zone (MBZ) and the extended Brillouin zone (EBZ) for $d=4$ stripes. 
The momenta in the EBZ label the $N$ plane wave states of the system.
Alternatively one can label the plane wave states by a  
reduced momentum $\tilde{\bf k}$ in the first MBZ (the central polygon
in Fig.~\ref{fig:pbz}) and a set of $N_a$ 
reciprocal superlattice vectors 
${\bf Q}\equiv m_1 {\bf b}_1 + m_2 {\bf b}_2 $ with $m_i$ integer,
the momentum of the plane wave state being given by 
${\bf Q}+\tilde{\bf k}$.
The dots in Fig.~\ref{fig:pbz} indicate a possible choice for this
set. Each dot defines also a higher MBZ's with the dot at the center.

Magnetic excitations are obtained by computing
random-phase approximation (RPA) like fluctuations on top of the GA
saddle-point\cite{sei01,sei04b} within the TDGA
which fulfills standard sum rules and on small
clusters yields excitation
spectra in very good  agreement with exact diagonalization.\cite{sei04b}

The basic ingredient of this approach is the interaction kernel. In the
spin channel this is obtained by
expanding Eq.~\ref{EGA} up to second order in the spin fluctuations 
\cite{sei04b}
\begin{eqnarray}
\delta E^{spin}&=&
\sum_{ij}V_{ij} \delta S_i^+\delta S_j^- \label{espin} \\
&+& \sum_{ij}M^\uparrow_{ij}[\delta S_i^+\delta\langle
c^\dagger_{i\downarrow} c_{j\uparrow}\rangle_0+\delta\langle
c^\dagger_{j\uparrow} c_{i\downarrow}\rangle_0\delta S_i^-]\nonumber
\\ &+& \sum_{ij}M^\downarrow_{ij}[\delta S_i^+\delta\langle
c^\dagger_{j\downarrow} c_{i\uparrow}\rangle_0+\delta\langle
c^\dagger_{i\uparrow} c_{j\downarrow}\rangle_0\delta S_i^-].
\nonumber
\end{eqnarray}
Here we have defined the following interaction matrices
\begin{eqnarray}
V_{ii}&=&\sum_{j,\sigma} t_{ij} \lbrack \label{eq:vloc}
\langle c^\dagger_{i\sigma}c_{j\sigma}\rangle_0
+ \langle c^\dagger_{j\sigma}c_{i\sigma}\rangle_0 \rbrack
z^{0}_{j,\sigma,\sigma} \frac{\partial^2 z_{i,\sigma,\sigma}}
{\partial S^+_i\partial S^-_i} \\
V_{i\neq j}&=& t_{ij} \lbrack
\langle c^\dagger_{i\uparrow}c_{j\uparrow}\rangle_0
+ \langle c^\dagger_{j\downarrow}c_{i\downarrow}\rangle_0 \rbrack
\frac{\partial z_{i,\uparrow,\downarrow}}{\partial S^+_i}
\frac{\partial z_{i,\downarrow,\uparrow}}{\partial S^-_i} \\
M^\sigma_{ij}&=& t_{ij} z^{0}_{j,\sigma,\sigma}\frac{\partial
z_{i,\uparrow,\downarrow}}{\partial S^+_i}
\end{eqnarray}
Basically the TDGA is closely related to the time-dependent 
Hartree-Fock (HF) approximation which in the limit of small
oscillations corresponds to the RPA.\cite{noterpa}
The structure of the equations is the same but the interaction kernel
is substantially different. Indeed in the present approach there is a  
space dependency of the effective on-site term $V_{ii}$ 
(instead of the bare '$U$') 
which follows the modulation of the underlying saddle point solution
(cf. Fig. \ref{fig0}). In addition
the GA+RPA formalism generates effective interactions between spin
fluctuations on different sites ($\sim V_{i\neq j}$) and between
intersite spin flip processes ($\sim M^\sigma_{ij}$). However, the
strength of the latter interactions is in general much 
smaller than the on-site contribution $V_{ii}$. 

In order to evaluate the RPA series we first have to evaluate
the ('zeroth order') susceptibilities on the GA level, e.g. 
\begin{eqnarray}
&&\chi^{st,0}_{nm}(\tilde{\bf q},\omega) =\\
&&\frac{i}{N}
\int\!\! dt\,\,e^{-i\omega t}\langle{\cal T} A^s({\bf\tilde{q}+Q}_n,t)
A^t({\bf -\tilde{q}-Q}_m ,0)\rangle_{GA}.\nonumber
\end{eqnarray}
Here the vector 
$\lbrace A^s\rbrace$ is composed of the following operators:
$A_n^1 \equiv \hat S^+({\bf\tilde{q}+Q}_n)$ is the momentum representation of
the local spin fluctuation $S_i^+=c^\dagger_{i\uparrow} c_{i\downarrow}$.
$A_n^{1+\alpha} \equiv T^\alpha({\bf\tilde{q}+Q}_n)$ are the Fourier transformed 
of the intersite spin fluctuations 
$c^\dagger_{i\uparrow} c_{i+R\alpha\downarrow}$.
Both fluctuations are mixed by the term $M^\sigma_{ij}$ 
in Eq. (\ref{espin})
(cf. Ref. \onlinecite{sei04b} for details).
For a model with next- and next-nearest neighbor hopping we have
$\alpha=1,\dots,8$ orthogonal operators $T^\alpha({\bf\tilde{q}+Q}_n)$
so that together with $S^+({\bf\tilde{q}+Q}_n)$ the 
complete zeroth order correlation function 
$\chi^{st,0}_{nm}(\tilde{\bf q},\omega)$
is a $18d \times 18d$ matrix.
On the other hand we can also express the Fourier transformed 
interaction Eq.~(\ref{espin})
in terms of a $18d \times 18d$ dimensional matrix 
$\underline{\underline{\Gamma}}({\bf\tilde{q}})$ 
\begin{equation}
\delta E^{spin}= \sum_{stmn}A^s({\bf\tilde{q}+Q}_n)\Gamma^{st}_{nm}({\bf\tilde{q}})
A^t({\bf-\tilde{q}-Q}_m)
\end{equation}
so that formally the RPA summation can be written as
\begin{equation}
\underline{\underline{\chi}}({\bf\tilde{q}})
= \underline{\underline{\chi^0}}({\bf\tilde{q}}) -
\underline{\underline{\chi^0}}({\bf\tilde{q}}) \,\,
\underline{\underline{\Gamma}}({\bf\tilde{q}}) \,\,
\underline{\underline{\chi}}({\bf\tilde{q}}).
\end{equation}

Upon inverting this equation we finally obtain the transverse
spin correlation function as $\chi^\perp_{nm}({\bf\tilde{q}})
\equiv \chi^{11}_{nm}({\bf\tilde{q}})$.

\subsection{Spin-wave theory calculations}
In order to interpret the GA+RPA results we have made a LSWT computation of
the dispersion similar to those performed in Refs. \onlinecite{kru03,car04}.
For simplicity we take a model with only AFM nearest neighbor magnetic 
 interactions $J$ equal everywhere except for the ferromagnetic bonds
 where  the interaction has opposite sign (to ensure stability) and
 magnitude $J_F$ as shown schematically in Fig.~\ref{fig:ucell}.

\subsection{Dynamical structure factor}
Neutron scattering experiments probe the longitudinal and transverse
components of the dynamical structure factor 
$S^{\alpha\beta}(\bm{q},\omega)$ weighted by
polarization factors:\cite{lov84,lor05} 
 \begin{eqnarray}
 \label{eq:seff}
  S^{eff}(\bm{q},\omega)&=&
 \sum_{\alpha} \langle (1-\hat q_\alpha^2)\rangle_{dom}
    S^{\alpha\alpha}(\bm{q},\omega)\\ \nonumber
&=&\eta_\parallel S^{\parallel}(\bm{q},\omega)+\eta_\perp S^{\perp}(\bm{q},\omega)
\end{eqnarray}
where  ${\bm  q}\equiv {\bm k}-{\bm k}'$ 
($\hat {\bm q}\equiv {\bf q}/|{\bm q}|$), ${\bm k}$ (${\bm k}'$) 
is the initial  (final) wave vector of neutrons and 
an average over the  orientation of domains, $\langle
 ...\rangle_{dom}$ has been done. In Eq.~(\ref{eq:seff}) we have 
 defined the longitudinal ($\parallel$) and transverse contributions
 ($\perp$) with respect to the ordered magnetic moment.

Unpolarized experiments are sensitive to relatively sharp
features which are expected to correspond mainly to transverse 
fluctuations\cite{lor05}
therefore in the present study we restrict to the latter 
$S^{eff}\approx \eta_\perp S^{\perp}$. The average
polarization factor  $\eta_\perp$ depends on the
orientation of the magnetic domains. For a distribution  
of domains isotropic in spin-space  $\eta_\perp=2/3$. 
For a non isotropic distribution  $\eta_\perp$ depends also on the
scattering geometry with the minimum value  $\eta_\perp=1/2$.\cite{lor05}

In terms of the reduced momentum ${\bf\tilde{q}}$ and the reciprocal 
superlattice vector ${\bf Q}_n$ defined in the previous subsection the 
structure factor is given by,
\begin{equation}\label{strfac}
S^{\perp}({\bf\tilde{q}+Q}_n,\omega)= \frac{(g\mu_B)^2}{\pi} Z_d \int\!\! dt\,\,
\exp(-i\omega t) \mbox{Im}\chi^{\perp}_{nn}({\bf\tilde{q}},t) .
\end{equation}
Here we have included an intensity renormalization $Z_d$ similar
to LSWT.  For a half-filled system with a SDW and in the limit of
large $U$, GA+RPA, HF+RPA and LSWT give the same results. The latter is
well known to overestimate the intensities of the spin-wave modes  
roughly by a factor of two and the same problem arises in the RPA
approaches.   GA+RPA, HF+RPA interpolate smoothly between this strong 
coupling limit and the noninteracting case, however, in the latter case 
$Z_d=1$. The value appropriate for cuprate parameters will be
intermediate between these two limits. We refer to 
Ref.~\onlinecite{lor05} where  $Z_d$ and $\eta_\perp$  are 
discussed in detail.

$S^{\perp}({\bf q},\omega)$ in general acquires 
different weight in  MBZ's centered on different ${\bf Q}_n$ (cf. Fig.~\ref{fig:pbz}).
Since we are not considering the magnetic form factor, 
$S^{\perp}({\bf q},\omega)$ repeats with the
periodicity of the EBZ reciprocal lattice vectors $(0,2\pi)$ and
$(2\pi,0)$. 

The stripes correspond to a  long-range magnetically  ordered state which
within our RPA-like scheme induces a Goldstone mode going to zero frequency at 
momenta ${\bf\tilde{q}}=0$. However, 
this mode has negligible weight at 
${\bf q}\sim 0$ but acquires a significant strength in the higher magnetic
zones at ${\bf\tilde{q}+Q}$ with the magnetic reciprocal lattice
vector ${\bf Q}\ne 0$.
The largest spectral  weight, as shown below,
is located at the dots closest to $(\pm\pi,\pm\pi)$ in
Fig.~\ref{fig:pbz}. These points are referred to 
 as the incommensurate wave vector ${\bf Q}_s$.

\section{Parameters and saddle-point solutions}\label{parameters}

With respect to our previous investigation \cite{sei05} we have
significantly improved the resolution in frequency and momentum
space which allows us to consider up to $10.000$ lattice sites
and frequency intervals of $\sim 1$meV as discussed in the preceding
section.
Within the present one-band description we fix parameters by
the following two conditions.

1.) The value of the nearest neighbor hopping has been
  fixed to $t'/t=-0.2$ according to Refs.~\onlinecite{sei04a,pav01}.
In Ref.~\onlinecite{sei04a}
we found that the stripe filling is very sensitive to  $t'/t$ and 
it turned out that a ratio of $t'/t=-0.2$ is necessary 
to describe half-filled stripes as inferred from the 
incommensurability.\cite{yam98} This value is close to
the value obtained in a local density approximation study.\cite{pav01} 

2.) The spin-wave dispersion of the undoped system along the
magnetic zone boundary poses a rather stringent constraint on the
value of $U$. In fact, LSWT for the Heisenberg model with only
nearest-neighbor exchange interactions $J =4 t^2/U$ does not
yield any dispersion at all. Only higher order in $t/U$ exchange
contributions (cyclic exchange being the most important among
them\cite{taka77,roger89,lor99,col01}) start to produce a dispersion in the 
magnetic excitations along the magnetic zone boundary which 
is therefore very sensitive to the value of $U$.

\begin{figure}[bp]
\includegraphics[width=8cm,clip=true]{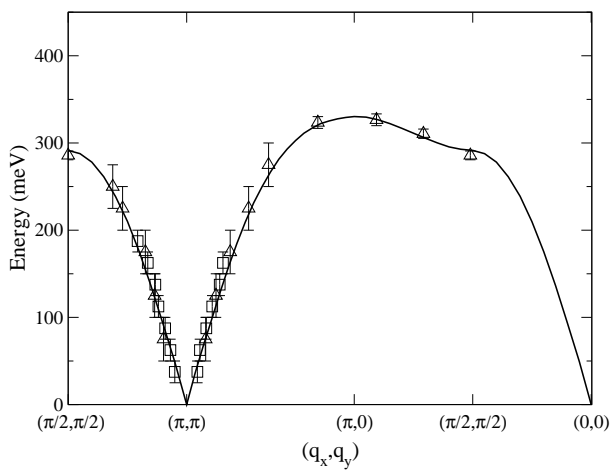}
\caption{Energy and wave vector dependence of magnetic
excitations in the half-filled system as obtained from
the GA+RPA approach for $U/t=7.5$, $t'/t=-0.2$ and $t=342$meV. Squares
and triangles correspond to data points from INS experiments at $T=10 K$ 
on La$_2$CuO$_4$ by Coldea et al. \protect\cite{col01}}
\label{fig1}
\end{figure}

We have computed the magnon spectra for the
half-filled system within the GA+RPA approach and fitted the
spin-wave dispersion of undoped La$_2$CuO$_4$ measured 
at $T=10 K$.\cite{col01}
As can be seen from Fig.~\ref{fig1} the parameter set $U/t=7.5$
yields  excellent agreement with the measured
dispersion also along the magnetic zone boundary. The fitting also
fixes the overall energy scale $t=342$meV.
This dispersion  is in slightly better agreement with experiment 
than a previous one\cite{lor05} with  $U/t=8$ which was
used in Ref.~\onlinecite{sei05} and enhances the already good agreement
between theory and experiment for the position of the resonance mode 
at $n_h=0.125$ discussed below. We note that for $U/t=10$ the fitting 
in the insulator is significantly worse.\cite{lor05}  
All our further calculations are without  free parameters.
It should be mentioned that a similar estimate for $U/t$ has been obtained from
quantum Monte-Carlo calculations of the magnon dispersion 
\cite{seng02} and is in the range $U/t\sim 6 , 8$ obtained from 
mapping the three band Hubbard model to the one-band
model.~\cite{sch92} Moreover, a similar value of  $U/t\sim 8$ has been 
obtained  in a recent  fit of the
doping dependence of spectral weights in optical data.~\cite{toh05}

Finally we note that although the present parameter set was optimized with
magnetic properties it also gives reasonable results for charge
excitations. Within GA+RPA one obtains a gap in the optical 
excitations of 1.8 eV which is close to the 2eV value of 
reflectivity experiments.\cite{uch91}

\begin{figure}[tbp]
\includegraphics[width=8cm,clip=true]{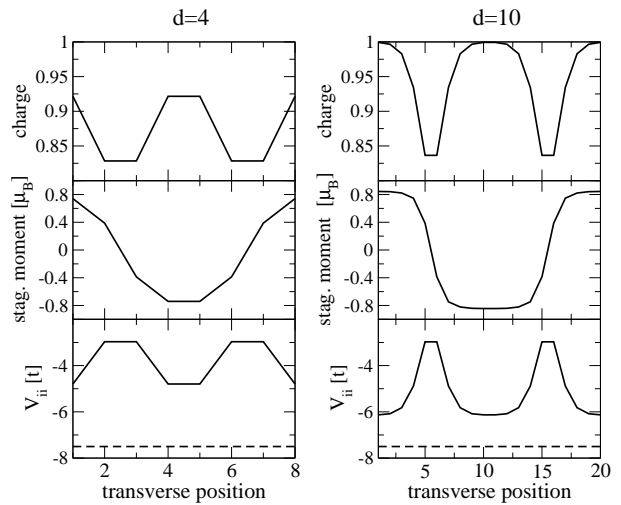}
\caption{Profile of the charge (top panels), staggered magnetic order
(middle panels) and effective interaction $V_{ii}$ (lowest panels) for
$d=4$ and $d=10$ bond-centered stripes in the GA.
 The dashed line in the
lowest panels indicates the 'bare' spin interaction of the HF+RPA  
$V_{ii}^{HF}=-U/t=-7.5$.}
\label{fig0}
\end{figure}

The saddle point solutions have been discussed in
Ref.~\onlinecite{sei04a}.
 Usually we consider stripe textures oriented
along the $y$ axis (vertical stripes) where site centered 
(SC) and bond centered (BC) solutions have similar energies. 
The magnetic unit cell for BC solutions is shown schematically in 
Fig.~\ref{fig:ucell}. Spin are ordered antiferromagnetically in
the vertical direction whereas the ferromagnetic spin alignment
between sites at transverse positions 
$0$ and $1$ indicates the presence of a
domain wall where also doped holes accumulate.  
In this paper we label stripes by the horizontal distance $d$ between the 
hole rich domain walls, the doping $n_h$ and the symmetry (BC or SC). 
In order to compare with experimental data in La$_{2-x}$Sr$_x$CuO$_4$
or La$_{2-x}$Ba$_x$CuO$_4$ we assume $n_h=x$.  

Within the present one-band model calculation we find that the more
favorable solutions at small doping are SC but with an energy difference
to BC stripes that is negligibly small in large systems.
This quasi-degeneracy is also found in other one-band calculations. 
\cite{whi98prl80,whi98prl81,fle00,fle01b}
On the other hand  BC textures turn out to be more stable  in the
more accurate three-band Hubbard model\cite{lor02b} and in first principle 
computations.\cite{ani04}  Therefore we pay special attention to BC
solutions at low dopings.  In the three-band model the quasi-degeneracy
appears at slightly larger doping and was related to the anomalous
optical properties.\cite{lor03}

In Fig.~\ref{fig0} we show a cut through the charge (top panels)
and staggered magnetic moment profile (central panels) for 
BC stripe solutions and stripe separation $d=4,10$ evaluated
within the GA. For distant stripes ($d=10$ in Fig.~\ref{fig0}) the regions
between the domain walls resemble the spin-density wave solution
of the undoped system (i.e. $n \approx 1$) whereas doped holes
progressively populate the AFM domains when stripes become closer.
The moment in the insulating regions is close to $0.8\mu_B$ and is reduced
with respect to the full value of $1\mu_B$ due to covalency.\cite{lor05} 
This is larger than the expected value in the insulator $\sim 0.6\mu_B$ 
because we are not including the feedback of the RPA fluctuations 
on the single-particle expectation values which, for the 
magnetic moment, is dominated by transverse fluctuations.\cite{sch89}

The weight of the magnetic Bragg peaks is determined by the Fourier transform
of the magnetic moment in the unit cell.\cite{tra96,lor05} From the more
intense Bragg peak one can define an ``ordered moment''. For LCO
codoped with Nd it was found\cite{tra96}  that this is about $0.1\mu_B$
whereas the present static GA approach yields $\sim 0.4\mu_B$.\cite{lor05} 
Apart from the lack of 
transverse fluctuations which may be even more relevant in the 
stripe phase it is likely that disorder, not taken into
account in our investigation, will depress the weight of Bragg peaks 
since they require coherence across many unit cells. In this regard it is
interesting that estimates from local probes, which are less sensitive
to disorder, are much 
larger and amount to ordered moments of $\sim 0.4\mu_B$.~\cite{tei00} 
This should  be compared with the average of the absolute value of the 
static moment $m^{GA} \approx 0.56\mu_B$ obtained within the GA. In
this case the overshoot is expected due to the omission of  
transverse fluctuations.  

Obviously the quality of our approach relies on the
presence of stripe correlations in the ground state of the
Hubbard model. The large clusters needed for clarifying this issue within exact
diagonalization methods presently render impossible a solution to the
problem. To estimate the quality of the GA with respect to more
advanced techniques (which are however restricted to smaller clusters
and specific boundaries) we have compared the charge distribution with
density matrix renormalization group (DMRG) calculations\cite{ws03} 
on a $7\times 6$ Hubbard cluster doped
with 4 holes and for $U/t=12$. This study also finds metallic
stripes.  Remarkably  the GA charge distribution (omitting RPA
fluctuations) is almost identical to the DMRG. 
In this case the RPA feedback effect is expected to be 
negligible in the insulator and small in the
stripe phase since the lowest energy excitations are gapped in the
former and limited to a small phase space in the latter.  

 The  similarity between the exact and the GA charge distribution 
gives further support to the quality of our starting point solutions
in contrast to HF theory where one finds a qualitatively wrong 
ground state for realistic parameters.\cite{sei04a} In this regard
GA+RPA becomes essential to proceed.

The local part of the effective interaction Eq.~(\ref{eq:vloc})
(lowest panel in Fig.~\ref{fig0}) closely follows the charge
modulation. This is in contrast to HF where  the effective interaction
between spin fluctuations is uniform and given by the bare value of $-U$. 
We find that the GA modulated interaction is the dominant effect
for obtaining a  
high-energy response in much closer agreement to experimental 
data, as described below, than the related HF+RPA computations.
~\cite{moe04,kane00,kane02} The interaction is also 
strongly renormalized in our approach, one of the reasons why GA+RPA 
performs much better than HF+RPA upon comparing with exact diagonalization
results.\cite{sei01} It is interesting to notice that the
effective value of the interaction we find is close to the value
usually taken for the bare interaction in HF+RPA computations ($U\approx 4t$) 
in order to obtain physically reasonable results.\cite{moe04,kane00,kane02} 
This phenomenological decrease of $U$ may compensate for some of the 
inadequacies of HF+RPA producing a theory more similar to GA+RPA.

\section{Results}\label{result}

\begin{figure}[tbp]
\includegraphics[width=8.5cm,clip=true]{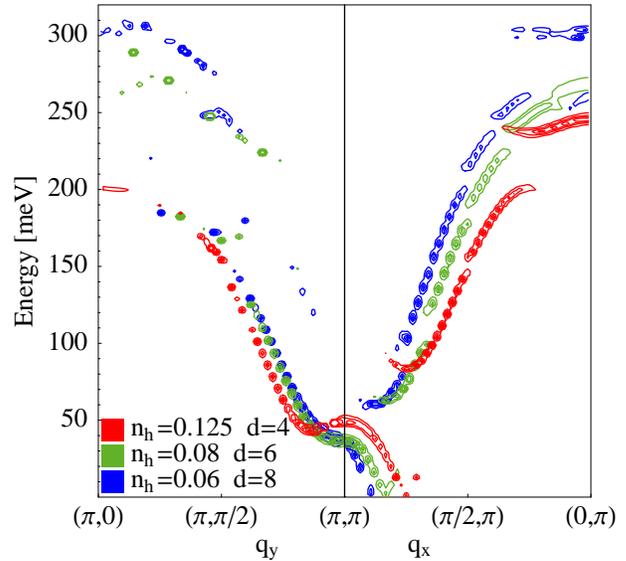}
\caption{(Color online) Dispersion of magnetic excitations along the cut
$(\pi,0)\to(\pi,\pi)$  and $(\pi,\pi)\to(0,\pi)$ for BC stripes and
 dopings $n_h\le 1/8$ and stripe separation 
 $d \ge 4$. The dispersion is obtained from 5 contours of 
$ \omega \chi^{''}_q(\omega)$ (which is dimensionless) at regular
 intervals between 0 and 7. 
} \label{fig:d8d4}
\end{figure}

\begin{figure}[tbp]
\includegraphics[width=8.5cm,clip=true]{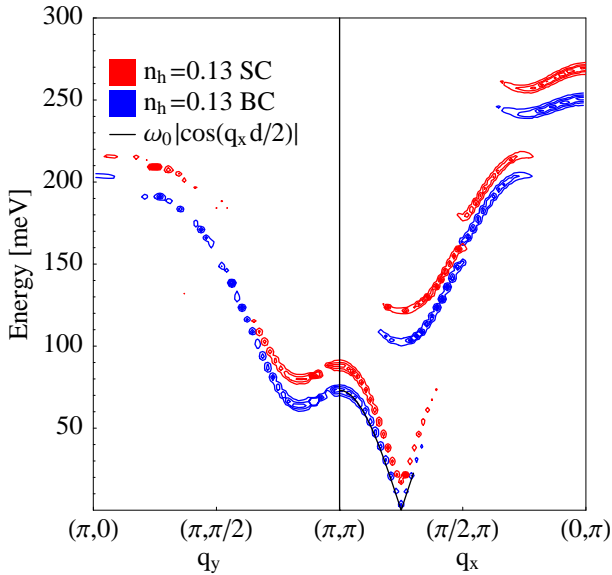}
\caption{(Color online) Dispersion of magnetic excitations along the cut
$(\pi,0)\to(\pi,\pi)$  and $(\pi,\pi)\to(0,\pi)$ for BC and SC stripes
and  $n_h=0.13$ and stripe separation is $d=4$. The SC data has been
shifted by 10meV for clarity. We also show the fit of the acoustic
branch with Eq.~(\ref{eq:acoustic}). The dispersion is obtained from 
5 contours of $ \omega \chi^{''}_q(\omega)$ at regular
 intervals between between 0 and 7.
} \label{fig:scbc.13}
\end{figure}

\begin{figure}[tbp]
\includegraphics[width=8.5cm,clip=true]{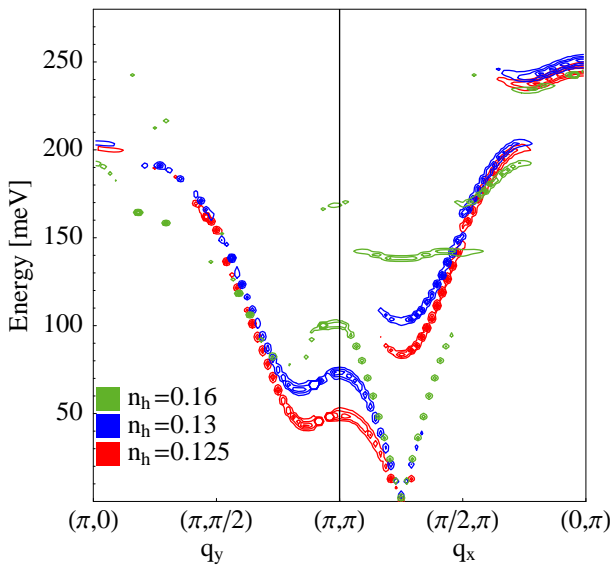}
\caption{(Color online) Dispersion of magnetic excitations along the cut
$(\pi,0)\to(\pi,\pi)$  and $(\pi,\pi)\to(0,\pi)$ for BC stripes at $d=4$
and different dopings. The dispersion is obtained from 5 contours of 
$ \omega \chi^{''}_q(\omega)$  at regular
 intervals between 0 and 7.
} \label{fig:d4}
\end{figure}

Figures~\ref{fig:d8d4},\ref{fig:scbc.13} and \ref{fig:d4} 
report magnetic excitations  for various dopings and stripe
separations. The spectra are for selected cuts parallel (right panels)  
and perpendicular  (left panels) to the stripes and correspond 
to intensity plots
of the imaginary part of the transverse magnetic susceptibility times
the frequency $ \omega \chi^{''}_q(\omega)$. The frequency
factor cancels a $1/\omega$ intensity divergence at the Goldstone mode
 and allows to visualize all energies with the same intensity scale.

The magnetic excitations for SC and BC  
stripes are almost identical with only minor 
differences regarding the intensity distribution and the dispersion 
as shown in Fig.~\ref{fig:scbc.13} for $n_h=0.13$. In the
other panels we therefore only report BC solutions which are the ones more
likely to be stable at low doping as discussed above.  
This similarity is in strong contrast to LSWT calculations for which the 
spectrum is substantially different.\cite{kru03,car04}
Indeed, within LSWT the number of 
bands is $d$ for BC stripes and $d-1$ for SC stripes. In the latter
case the core spins are assumed to be in $S=0$ states and thus to be
passive. In GA+RPA instead all electrons are active and the 
number of low-energy magnetic bands is $d$  irrespective of the 
symmetry of the stripe. (This is not obvious from the figures 
for large $d$ because the GA+RPA bands have small gaps and are only visible
in restricted parts of the Brillouin zone.)

As expected a Goldstone mode emerges from  the incommensurate wave
vector. In GA+RPA a small gap ($\sim 6$meV)
appears in some runs due to convergence problems of the solutions. 
In these cases we subtract the gap to obtain the physically correct 
gapless spectra. 

The Goldstone mode excitations correspond to long wave-length
transverse fluctuations of the magnetic order parameter. i.e. 
acoustic spin waves. The dispersion relation consists of a cone 
around the incommensurate wave
vector with an anisotropic intensity distribution. 
The intersection of the spin-wave cone with the $(q_x,\omega)$-plane
with $q_y=\pi$ gives rise to two acoustic 
branches emerging from the incommensurate wave vector ${\bf Q}_s$. 
The large intensity one disperses towards ${\bf Q}_{AF}=(\pi,\pi)$ where it
reaches a saddle point ($n_h<0.09$) or a local maximum 
($n_h>0.09$). We associate the energy at the AFM wave vector,
$\omega_0$, with the resonance energy.\cite{resonance}

\begin{figure}[tbp]
\includegraphics[width=8.5cm,clip=true,bb=88 4 376 200]{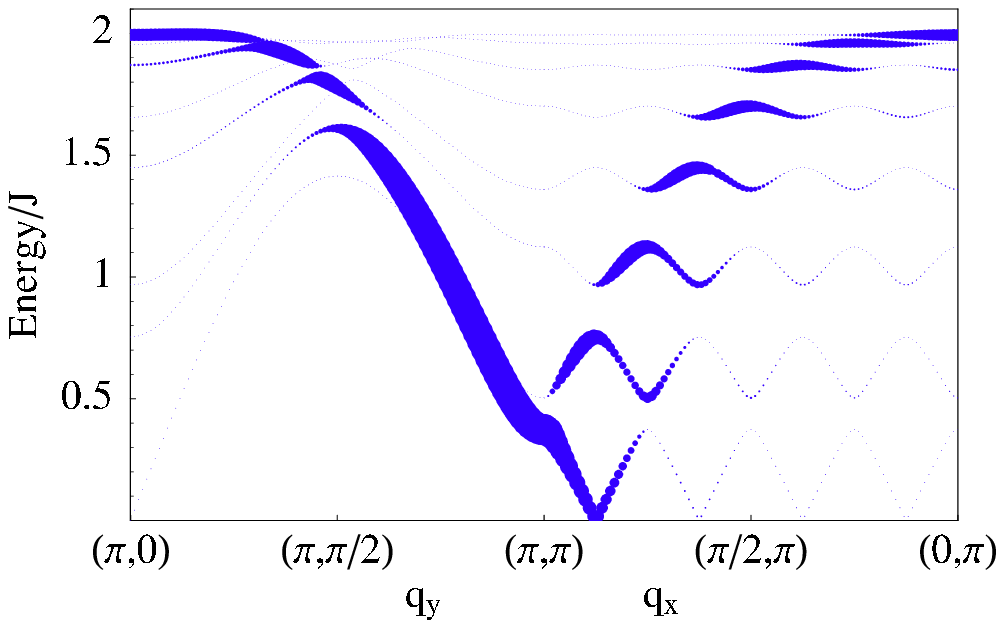}
\includegraphics[width=8.5cm,clip=true,bb=88 4 376 200]{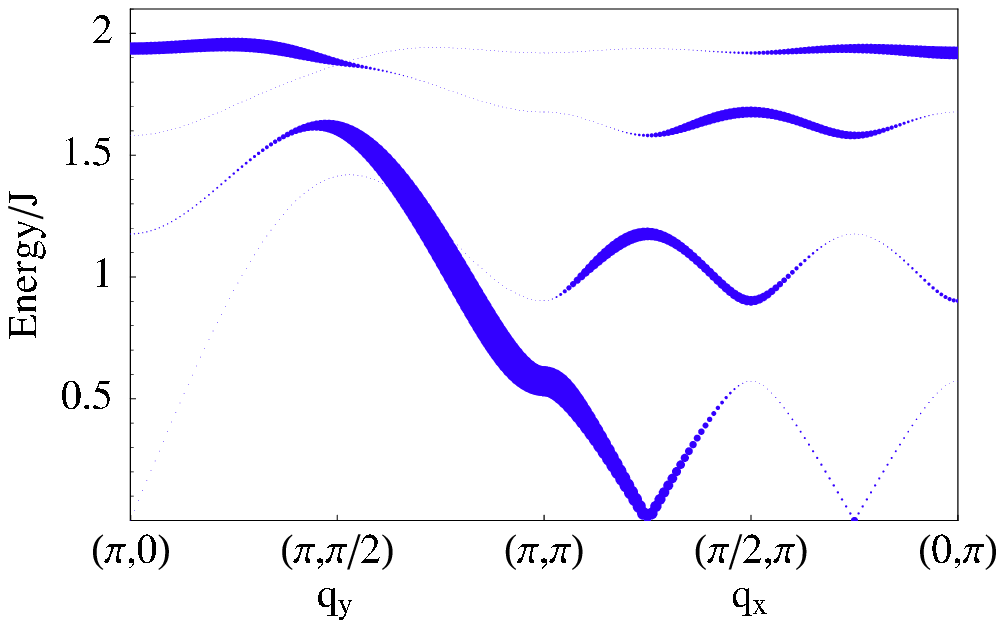}
\caption{Dispersion of magnetic excitations along the cut
$(\pi,0)\to(\pi,\pi)$  and $(\pi,\pi)\to(0,\pi)$ for BC stripes
oriented along the y-direction within LSWT for $d=8$ (top) and $d=4$
(bottom). The radius of the points are proportional to the 
intensity times the energy of the excitation. Modes with negligibly
weight are plotted with a small dot.    
We used a uniform value of the AFM interaction $J$ across all the bonds 
except for the ferromagnetic bonds where the sing is opposite to
ensure stability and the modulus is $J_F=0.2J$ as shown in Fig.~\ref{fig:ucell}. } \label{fig:swtp2}
\end{figure}

The branch emerging from ${\bf Q}_s$ in the cuts and dispersing 
outwards  (i.e. away from ${\bf Q}_{AF}$) rapidly looses 
intensity with increasing energy. As can be seen from 
Fig.~\ref{fig:d8d4} this effect is more pronounced for large stripe
separations. 
Indeed for $n_h=0.06$ the outwards dispersing branch is 
hardly visible.

In Fig.~\ref{fig:swtp2} we show the LSWT results for $d=8$
and $d=4$ BC stripes with $J_F=0.2$. We can reproduce some 
qualitative features of the GA+RPA
dispersion relation by taking such a small value of $J_F$ as discussed in
detail below. The intensity difference  effect between inwards
and outwards dispersing Goldstone mode is also observed in
the LSWT calculations (Fig.~\ref{fig:swtp2}), however, the difference
 remains  constant as a function of $d$ and is less
 apparent for large stripe separations.\cite{note1}

Since the outwards dispersing branch has not been
detected yet we cannot make a detailed comparison of the relative
intensities. Our GA+RPA results suggest that this branch should 
be difficult to observe in the  underdoped regime. 
 For example for $n_h=1/8$ 
the intrinsic momentum width of the inward branch observed in the
experiment will mask the small
low energy region where the  outward branch has significant 
weight (cf. Fig.~\ref{fig:d4tran}).

For $n_h=0.16$ we expect this problem to be less severe and the branch
should become observable, 
however, low energy INS data\cite{chr04}   do not
reveal any signature of the outwards dispersing branch. It may be that 
our theory overestimates the corresponding intensity in
this case or that damping effects, not taken into account in our approach, 
become dominant. More experimental and theoretical work
is needed to clarify this point. 
It should be mentioned in this regard that  HF+RPA calculations \cite{moe04}
suggest a further suppression of the intensity of the outwards dispersing 
modes from the inclusion of a $d$-wave order parameter.

The two GA+RPA acoustic branches along the $(q_x,q_y=\pi)$ direction
can be fitted by:
\begin{equation}
  \label{eq:acoustic}
\omega(q_x,\pi)=\omega_0|\cos(q_x d/2)|  
\end{equation}
as shown in Fig.~\ref{fig:scbc.13}. The corresponding velocity
perpendicular to the stripe is:
\begin{equation}
  \label{eq:cdq}
  c=\frac{\omega_0 d}2.
\end{equation}
The spin wave cone is only weakly anisotropic 
(cf. Figs.~\ref{fig:x008},~\ref{fig:x016})
so that the spin-wave velocity along
the stripe is not far from this estimate.

\begin{figure}[tbp]
\includegraphics[width=8.5cm,clip=true]{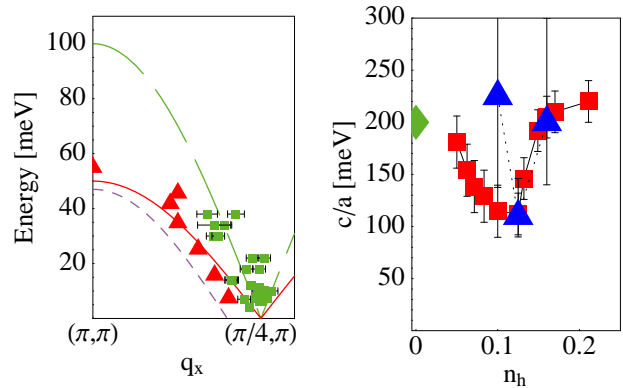}
\caption{(Color online) Left panel: 
Dispersion of magnetic excitations along the cut
$(\pi,\pi)\to(\pi/4,\pi)$ for BC stripes for dopings $n_h=0.1$ (short dashed
line),
0.125  (full line) and 0.16 (long dashed line). 
We include the experimental
data from Ref.~\onlinecite{tra04} for $n_h=0.125$ (triangles)  and 
from Ref.~\onlinecite{chr04} for  $n_h=0.16$ (squares).
 Right panel: velocity of the
collective mode $c$ divided the lattice spacing $a$. We report the
theoretical result (squares) and the experimental result
(triangles) both obtained by fitting Eq.~(\ref{eq:acoustic}) to the
data. The diamond at $n_h=0$ is the spin-wave velocity in the
insulator. } \label{fig:acoustic}
\end{figure}

In Fig.~\ref{fig:acoustic} we show available experimental data for
LBCO\cite{tra04} $n_h=0.125$  and LSCO \cite{chr04} $n_h=0.16$
together with the GA+RPA dispersion. There is excellent
agreement between theory and experiment for the position of the low-energy
acoustic branches and the velocity $c$ of the collective
mode. We caution, however, that the errors in the determination of the
experimental spin-wave velocity are rather large as shown by the error bars 
in the right panel of  Fig.~\ref{fig:acoustic}. Even a spin-wave
velocity independent of doping is compatible with the present experimental
data. To determine the theoretical spin-wave velocity, shown also in
the right panel, we used
Eqs.~(\ref{eq:acoustic}),(\ref{eq:cdq}). For some runs it was
necessary to include a correction to Eq.~(\ref{eq:acoustic}) of the
form $\omega_1 |\sin(q_x d)|$ with $\omega_1$ in the range $\pm 6$meV.
This correction, however, was too noisy due to convergence problems of the
runs so we include this effect as error bars in the theoretical data.

The theoretical velocity as a function of doping  has a pronounced minimum
at $n_h=1/8$ which seems to be supported  by the available experimental
data with the mentioned caution on the errors involved.  

Inspection of  Eq.~(\ref{eq:cdq}) reveals that at small concentrations 
the velocity is mainly determined by $d\sim 1/n_h$ due to the weak doping
dependence of $\omega_0$ in this regime [cf. Fig. \ref{fig3}(a)].
However, for $n_h > 1/8$ the stripe periodicity stays constant ($d=4$)
and the rising of $\omega_0$ (cf. Fig. \ref{fig3}a) causes  the increase of the velocity and
thus the appearance of the minimum at $n_h=1/8$.
For $n_h=0.1$ the available data for the collective 
mode\cite{chr04} (not shown) imply a velocity substantially 
larger than what we found (as shown in the right panel of
Fig.~\ref{fig:acoustic}). The predicted dispersion is shown in the
left panel of  
Fig.~\ref{fig:acoustic} with the dashed line. We expect that the
saddle point energy $\omega_0$ is similar or slightly smaller 
than at $n_h=0.125$ whereas the present low-energy experimental
data\cite{chr04} extrapolate to a significantly larger energy. 
A high-energy experiment should clarify the situation on this point.

For $n_h>0.09$ the GA+RPA computation additionally shows a ``roton
like'' minimum in the direction of the stripe 
(Figs.~\ref{fig:d8d4},\ref{fig:scbc.13},\ref{fig:d4}).
 This effect cannot be explained in LSWT without invoking longer range
interactions. 

The minimum is located close to $(\pi,3\pi/4)$ which is
equal to ${\bf Q}_s$ with $x$ and $y$ directions interchanged.
It indicates a tendency of the stripe to develop a 
spin-density wave with period $8a$ along the stripe.
One can speculate that for different parameters  
the system may develop such an instability 
resulting in a ground state with a checkerboard 
pattern as observed on the surface of some compounds.\cite{han04}

The maximum at  
${\bf Q}_{AF}=(\pi,\pi)$ and the roton minimum give rise to a peak 
in the momentum integrated structure factor which is slightly 
lower in energy than the saddle point ``resonance'' 
at the AFM wave vector (Fig.~\ref{fig3.5}). 
Especially at low doping (i.e. large stripe separations) the
excitations around the $(\pi,\pi)$ region carry significant weight
which in the integrated spectrum is further enhanced due to the
saddle-point structure.

\begin{figure}[tbp]
\includegraphics[width=8.5cm,clip=true]{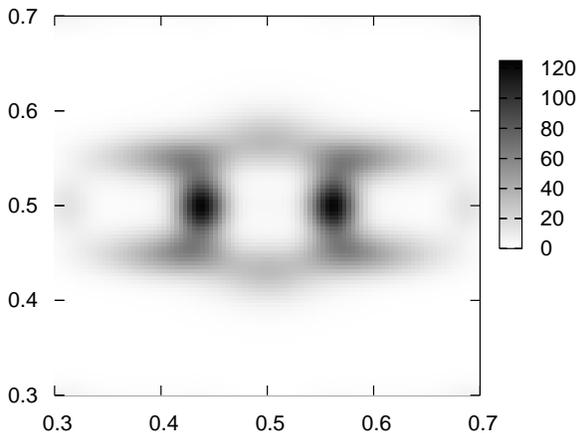}
\caption{Contour plot of $\chi^{''}_q(\omega)$ as a function of wave
  vector for 
$\omega=45.6$meV and $n_h=0.125$. 
The plot range for the axes is the
 same as in Fig.~1(c) of Ref.~\onlinecite{hin04}.
 Wave vectors are in  units of $2\pi/a$. 
}
\label{fig:con42}
\end{figure}

Because of the roton minimum a constant cut at an energy moderately below the
resonance shows intensity both along and perpendicular to the direction 
of the stripe as shown in
Fig.~\ref{fig:con42}. Although our study focuses on LCO it is
remarkable that a similar effect has been recently
observed in an untwinned sample array of YBCO by 
Hinkov and coworkers.\cite{hin04} 
The intensity difference between the two directions in also in
qualitative agreement with our result.

In Ref.~\onlinecite{hin04} this quasi two-dimensional distribution
of intensity is interpreted as evidence against a ``rigid stripe array''
which is expected to produce a more anisotropic pattern. It is clear
that Ref.~\onlinecite{hin04} cannot be considered as evidence against
an {\it oriented}  stripe array since these authors definitively found 
an asymmetry between $x$ and $y$ directions, i.e. a breaking 
of $C_4$ lattice symmetry.
As already outlined in the introduction the latter
can be viewed as the fingerprint 
of an oriented stripe ground state (in the sense of a 'stripe nematic
phase' \cite{kiv03}) but 
may also be due to a structural anisotropy as proposed 
in Ref.~\onlinecite{erem205}.
Without further quantitative estimates one cannot distinguish between
both scenarios from Ref.~\onlinecite{hin04}.

Our result shows that long-range order with 
oriented stripes is compatible with a quasi two-dimensional distribution
of intensity between the saddle-point energy and the roton minimum.
At lower energies we predict a strongly anisotropic intensity as long
as the dispersion is not cutoff by the spin gap. 
Can we agree with the conclusions of Hinkov and coworkers? This
becomes a question of semantics.
Our stripes are quite different from the ``rigid stripe array''
assumed in LSWT since our RPA computation is performed
on top of a much softer modulation and also has a much larger phase 
space for fluctuations which substantially change the results.
Interpreting  ``rigid stripe array'' in the LSWT 
sense we can agree with the statement Ref.~\onlinecite{hin04} 
since, as we have shown, LSWT in its simpler form is not able 
to reproduce this result.

We believe that the fact that our results 
qualitatively reproduce the observed
two-dimensional intensity map observed by Hinkov {\it et al.} 
is  evidence for the
existence of the roton minimum which was obtained\cite{sei05} 
before the data of Ref.~\onlinecite{hin04} became available. 
Naturally, a quantitative comparison should include 
consideration of the YBCO bilayer structure which would
modify the energy of the resonance and may affect intensities.
In addition, a description which allows for fluctuating stripes from
the start\cite{voj05} 
may be more appropriate for YBCO materials since it can capture the
appearance of a spin gap. 
In this regard some details of our map differ from the experimental
data (round shape, external branches) perhaps because of the above
reasons.  One should keep in mind, however, that the 
map of Hinkov {\it et al.}\cite{hin04} is obtained on the basis of a
simple model fitted to 4 scans in each direction of which only 3
intersect the ring. Clearly details not considered by their model will
not show up. This should be compared with a
typical experiment at ISIS like Ref.~\onlinecite{tra04,hay04} where
the maps involve hundreds of scans per direction.

In order to compare with experimental data in twinned samples or for 
systems in which stripes are stacked 
orthogonal to each other from one plane to the next, one should
average over the two possible orientations of the stripes (vertical and
horizontal). 
Whereas the high-energy magnetic excitations in the direction of
the stripes evolve from the saddle-point
at ${\bf Q}_{AF}$ the spectrum perpendicular to the stripes is composed
of several optical branches which are not directly connected to 
the 'resonance'. Nevertheless it is interesting to observe that in GA+RPA
the high-energy dispersions parallel and
perpendicular to the stripes (Fig.~\ref{fig:d4tran}) are rather close 
indicating an almost isotropic magnetic response at large frequencies.
This high energy 'isotropy' is only present up to $n_h \sim 0.15$
where due to the interaction between magnetic domains the optical
branches perpendicular to the stripes start to develop a significant 
gap structure.  
Fig.~\ref{fig:d4tran} reveals that also for the optical 
branches our calculation yields reasonable agreement with the 
available data.\cite{tra04} In striking contrast  
LSWT (Fig.~\ref{fig:swtp2}) predicts a series of optical branches in the 
direction perpendicular to the stripe that are not observed and an
anisotropic high energy response.  
The GA+RPA high energy excitations shown in Fig.~\ref{fig:d4tran} are
slightly displaced towards higher momenta. However, we find excellent
agreement for the slope which defines an effective 
high-energy spin-wave velocity. 
The shift of our excitations towards
 higher momenta may be due to the use of the oversimplified one-band
 Hubbard model. It would be interesting to see how the high energy
 part of the dispersion appears in
 the three-band Hubbard model. 

A quantitative fit to the X-type
dispersion shown in Fig. ~\ref{fig:d4tran} can also be achieved
with coupled spin-leg ladder models by fine tuning the interstripe
coupling to a quantum critical point.\cite{uhrmag}
These theories, however, rely on an even charge periodicity 
of the stripes (as for  $n_h=1/8$). 
In contrast, we obtain qualitatively similar spectra for  
$n_h=1/8$ ($d=4$) and $n_h=1/10$ ($d=5$)  
in accord with experiment in LCO\cite{tra04,chr04}. On the other hand our
approach cannot be easily generalized to incorporate
spin gapped states as in YBCO which are naturally
explained in a ladder picture.\cite{voj04,uhr04,uhr05,uhrmag}
Another important difference of the ladder scenario with respect to the 
present work is that at high energies the ladder computation finds
propagating modes only in the direction of the stripe whereas our calculation  
yields more isotropic spectra in agreement with recent 
experiments.\cite{hin06}

\begin{figure}[tbp]
\includegraphics[width=8.5cm,clip=true,bb=88 4 376 292]{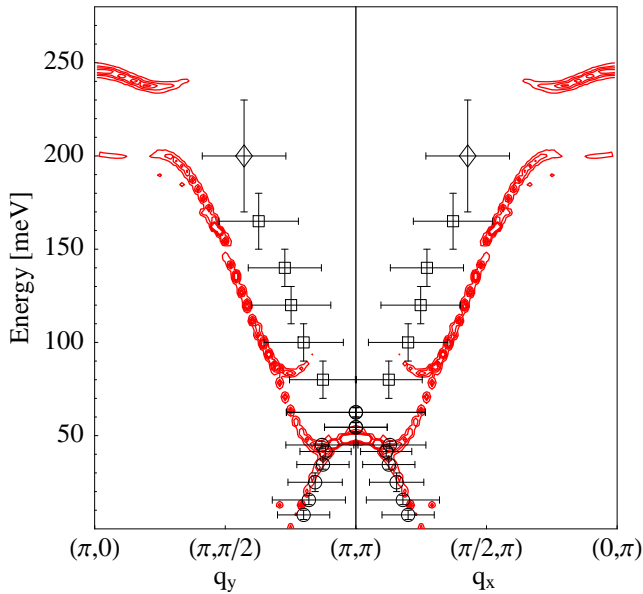}
\caption{(Color online) Dispersion of magnetic excitations along the cut
$(\pi,0)\to(\pi,\pi)$  and $(\pi,\pi)\to(0,\pi)$ for BC stripes at
$d=4$ and $n_h=0.125$. We have averaged the data over the two
possible orientations of the stripe, vertical and horizontal. 
The dispersion is obtained from 5 contours of 
$ \omega \chi^{''}_q(\omega)$ at regular intervals between 0 and 10. 
We also show the experimental data 
from Ref.~\onlinecite{tra04} (courtesy of J. Tranquada). 
For the latter horizontal ``error'' bars 
indicates the half width of the peaks whereas vertical  ``error'' bars
indicate an energy interval of data integration. 
} \label{fig:d4tran}
\end{figure}

Constant energy cuts of the dispersion change shape as the energy of
the cuts is increased. We have already shown\cite{sei05} that our data closely
resemble the changes seen in the experiment at $n_h=1/8$.\cite{tra04}
In Figs. \ref{fig:x008},\ref{fig:x016} we show contour plots of the
transverse susceptibility for a strongly underdoped ($n_h=0.08$)
and an almost optimally doped ($n_h=0.16$) system averaged over the
two possible orientations of the stripes.
In the first case we consider an underlying $d=6$ BC stripe solution
which yields a saddle-point energy at $\omega_0=43 $meV.
Below this energy the 'resonance feature' splits into the four 
incommensurate Goldstone modes (from the superposition of vertical and
horizontal stripes). The associated spin-wave cones display a strongly 
asymmetric intensity  distribution where the weight is mainly confined
in the parts closest to ${\bf Q}_{AF}$ in accord with the intensity difference
of the acoustic branches shown in Fig.~\ref{fig:d8d4} and discussed above. 

Above $\omega_0$ the
intensity spreads into a ring shaped feature which increases with
energy. The strongest scattering is on the corners of  a square at 15meV 
and rotates by
45$^\circ$ at 60meV resembling the behavior found at $n_h=1/8$ in
theory\cite{sei05} and experiment\cite{tra04}.

The case of $d=4$ BC stripes at doping $n_h=0.16$ is shown in
Fig. \ref{fig:x016}. The saddle-point energy is at $\omega_0=102 $meV.  
At low energies the spin-wave cone displays significant weight in all 
directions which leads to the pronounced intensity of the outwards 
dispersing branch as discussed already in context of Fig. \ref{fig:d4}.
The complex structure at $87 $meV is composed of the incommensurate 
Goldstone modes and the branch dispersing from ${\bf Q}_{AF}$ towards the roton 
minimum which 
constitute the smaller ring around ${\bf Q}_{AF}$. Additionally one can observe 
a larger ring which intensity comes from the branch dispersing away from
the roton minimum along the stripe direction. A weaker feature along
the diagonals originates from the outwards dispersing branches.
Above the saddle-point energy the intensity distribution again forms a
ring which increases in diameter with increasing energy.

Fig.~\ref{fig3} displays the doping dependence of the resonance
frequency and intensity for both BC and SC stripes evaluated within GA+RPA. 
In the strongly underdoped regime ($0.05 \le n_h \le 1/8$) we find an
approximately  linear relation $\hbar \omega_0 \approx \alpha
+ \beta n_h$  with $\beta \approx 250$meV and $\alpha \approx$
20 (30) meV for BC (SC) stripes in the GA+RPA. Note that for infinitely
distant stripes one expects $\omega_0 \to 0$ so that the
offset energy $\alpha$ is an artefact of the linear approximation
in the considered doping range.  A qualitatively similar linear 
behavior has been found in Ref.~\onlinecite{bat01}. 

\begin{figure}[tbp]
\includegraphics[width=8.5cm,clip=true]{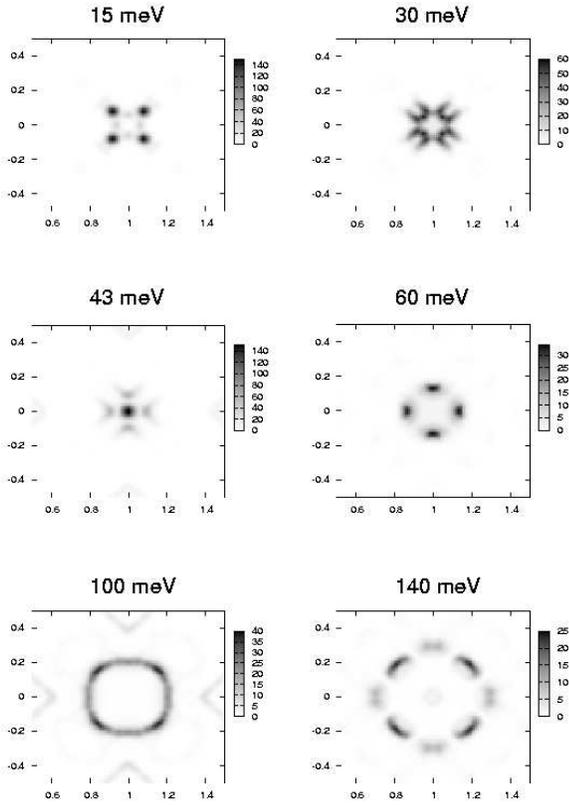}
\caption{Contour plots of $\chi^{''}_q(\omega)$ as a function of
  wave vector at
selected energies for doping $n_h=0.08$. The underlying GA solution is a
$d=6$ BC stripe array. The wave vector coordinate system has been rotated by
$45^\circ$ in such a way that the magnetic Brillouin zone
of an antiferromagnet occupies all the window of the plot  
with ${\bf Q}_{AF}=(\pi,\pi)$ at the center. Wave vectors are in 
units of $2\pi/(\sqrt(2)a)$. }
\label{fig:x008}
\end{figure}

\begin{figure}[tbp]
\includegraphics[width=8.5cm,clip=true]{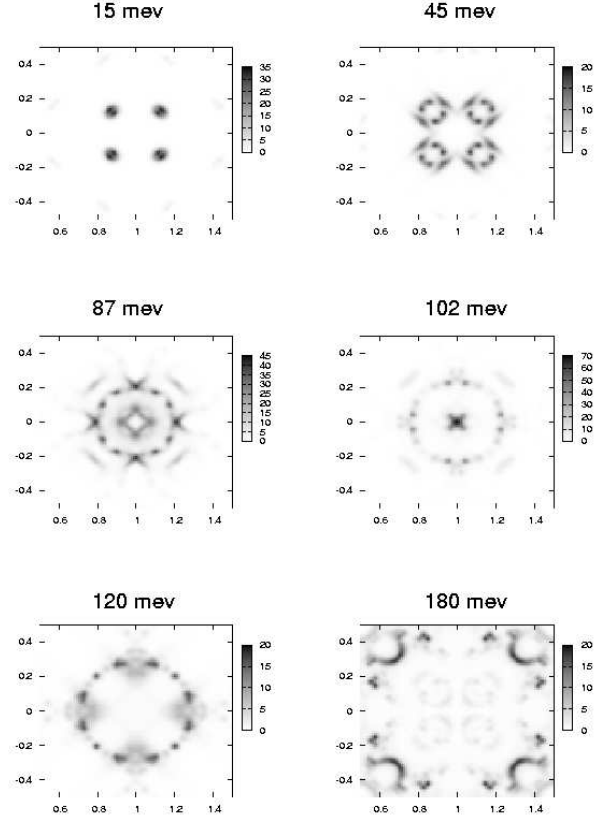}
\caption{Contour plots of $\chi^{''}_q(\omega)$ as function of
  wave vector at
selected energies for doping $n_h=0.16$. The underlying GA solution is a
$d=6$ BC stripe array. The wave vectors units are as in Fig.~\ref{fig:x008}.
}
\label{fig:x016}
\end{figure}

We also show in Fig.~\ref{fig3}(a) (full squares) the doping dependence of 
the resonance
energy  evaluated within LSWT. We assumed that the doping is
given by $n_h=1/(2d)$ which is valid for half-filled stripes with $n_h<1/8$.  
The energy of the resonance scales in
a similar way in both GA+RPA and LSWT for $n_h<1/8$ despite the fact that
we take the values of $J$ and $J_F$ as doping independent.
This is not surprising since for $n_h<1/8$ the charge profile of the
stripes do not overlap and doping proceeds by varying the distance
among the stripes keeping the electronic structure on the core of the
stripe essentially unchanged.\cite{lor02b,sei04a} 
 Therefore in this respect it makes little difference 
at low energy to consider the stripes as localized objects as in 
LSWT with a fixed value of $J_F$ or as more extended objects as in GA+RPA. 
It is interesting that the effective value of $J_F$ in order to
have an overall agreement with the GA+RPA results and with the
experiment is quite small as found in other LSWT studies.~\cite{kru03}

For $n_h>1/8$ effects due to the overlap of the stripe
cores become evident. In our previous computations\cite{sei04a,lor02b}
 within the GA we have found that in this regime it is energetically 
favorable to add holes into the $d=4$ stripe structure 
consistent with the doping independent incommensurability
seen in INS experiments.\cite{yam98} For both BC and SC
stripes this leads to a strong increase of the resonance energy
and therefore to a deviation from the linear
$\omega_0$ vs. $n_h$ behavior at low doping. An alternative scenario 
is phase separation as explained below.

\begin{figure}[tbp]
\includegraphics[width=8.5cm,clip=true]{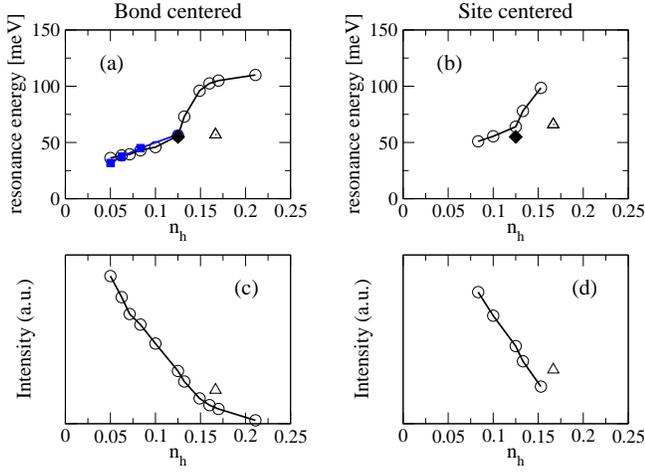}
\caption{(color online)
Doping dependence of the resonance frequency and
intensity  for BC (left panels) and SC (right panels) stripes within
the GA+RPA. The circles for $n_h>1/8$ correspond to $d=4$ solutions
whereas the triangle corresponds to the $d=3$ solution.  
The full diamond symbol is the experimental result from
Ref.~\protect\onlinecite{tra04}. The full squares in (a) 
are the LSWT result using $J_F=0.2J$ and $J=100$meV. } \label{fig3}
\end{figure}

\begin{figure}[tbp]
 \hspace*{-1cm}\includegraphics[width=8cm,clip=true]{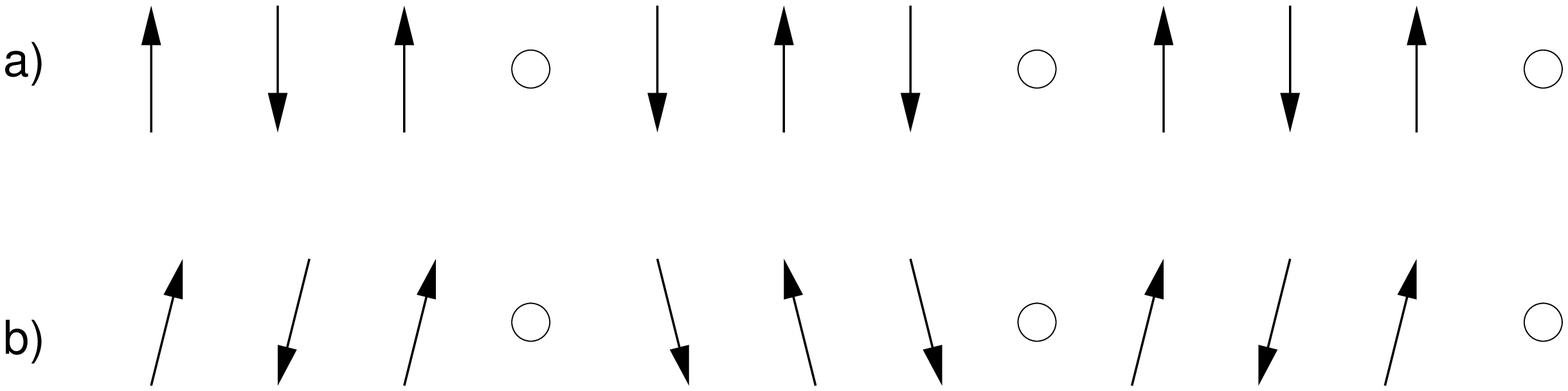}
 \caption{ a) A schematic cut perpendicular to the stripe 
through the N{\'e}el-type spin structure of
 $d=4$ SC stripes. b) Resulting spin structure upon adding
 transverse spin components $S_{\perp}\sim \cos({\bf Q R})$ with
 ${\bf Q}=(\pi,\pi)$} \label{fig5}
 \end{figure}

\begin{figure}[tbp]
\hspace*{-1cm}\includegraphics[width=6cm,clip=true]{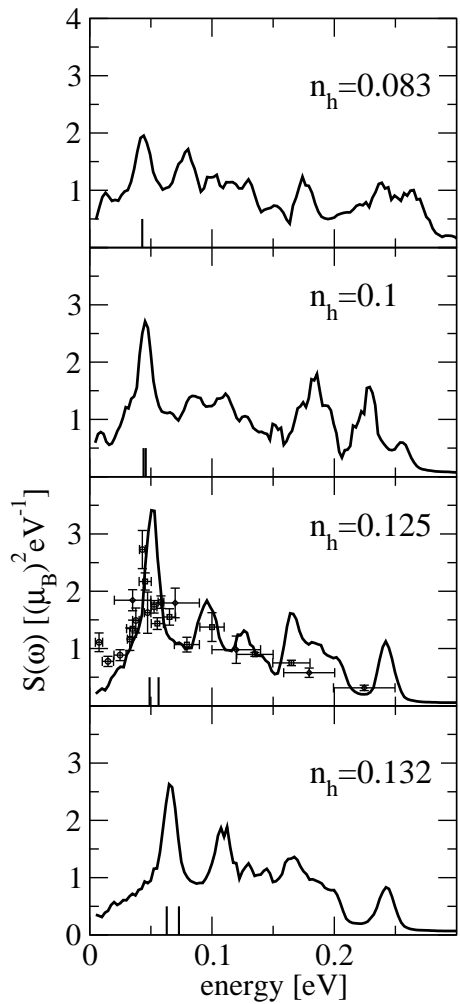}
\caption{Density of magnetic excitations $S(\omega)$ (see text)
for three doping levels $n_h=0.083, 0.1, 0.125, 0.132$. Theoretical curves
include a factor $Z_d \eta_\perp = 0.25$. The vertical bars indicate the energies of
the saddle-point at ${\bf Q}_{AF}$ and the roton-type minimum in the dispersion
along the stripes. Experimental data for
$n_h=0.125$ are from Ref.~\protect\onlinecite{tra04}.}
\label{fig3.5}
\end{figure}

The resonance mode is strongly influenced by the effective 
magnetic interactions across the domain wall. For example, in case
of a SC stripe depicted in  Fig.~\ref{fig5} the spins adjacent to the
core site are bridged by an effective AFM interaction $J'$ as it is
assumed in LSWT models.
The energetics of the resonance mode can be easily 
understood  by adding a small transverse component with momentum 
${\bf  Q}=(\pi,\pi)$ to the domain wall structure which provides a
simplified picture for the quasi-classical displacements of the spin 
in the resonance mode as shown in Fig.~\ref{fig5}(b). The spins in the  AFM regions keep
their relative angle constant whereas the angle formed by the 
spins across the domain wall strongly fluctuates, giving the dominant
contribution to the energy of order $J'$ . A similar picture holds 
for BC stripes with $J_F$ at the place of $J'$. 
(In reality the resonance mode is not
completely localized in the domain wall, as suggested by
Fig.~\ref{fig5}, which makes the dependence on $J'$ of $J_F$ 
more complicated.) 

The strong increase of the resonance energy
with doping for $n_h\ge 1/8$ [Fig.~\ref{fig3}(a)] can 
be understood if the effective  value of  
$J_F$ ($J'$) also increases across the BC (SC) domain walls. 
As shown in the Appendix~\ref{ap1}
 these couplings are dominated by transverse hopping processes of doped
holes. The increase of the stripe 
filling from $\nu=1/2$ for $n_h\ge 1/8$ induces a
concomitant enhancement of $J_F$ ($J'$) explaining the behavior of the
upper panels of Fig.~\ref{fig3}.

Since  upon doping the saddle point energy for $n_h < 1/8$ increases 
more slowly than the experimental $T_c$ 
our calculations predict a non-constant ratio between
$\omega_0$ and $kT_c$ in the underdoped regime of LCO. 
This ratio decreases from $\omega_0/kT_c \approx 27$ for
$n_h=0.05$ to $\omega_0/kT_c \approx 17$ for $n_h=1/8$.
This is significantly larger than the resonance energy to  $kT_c$
ratio 
$\sim 5.4$ as obtained for YBCO. 
Therefore if the saddle point excitation is
important for the superconducting mechanism its effect on $T_c$ is 
modulated by other factors.

In LCO materials the only experimental value for $\omega_0$
available up to now \cite{tra04} (shown as the full diamond in
Fig.~\ref{fig3}) falls exactly on the curve for BC stripes whereas
it is slightly below the $\omega_0(1/8)$ result for SC ones.
Although the difference is too small to be conclusive (given the
approximations involved) the fact that BC stripes give the best fit 
is in agreement with our previous finding within the
three-band model \cite{lor02b} that BC stripes are the more stable
textures in the underdoped regime of LCO cuprates.

One can obtain two more points for the resonance energy by 
extrapolating the experimental data
of Ref.~\onlinecite{chr04} with the aid of
Eq.~(\ref{eq:acoustic}). One obtains a similar result
as in the right panel of Fig.~\ref{fig:acoustic}. The extrapolated
experimental resonance is in good agreement with theory for 
$n_h=0.16$ whereas it is at much higher energies for $n_h=0.1$.

Around optimal doping one finds a
proliferation of mean-field solutions with energy similar to stripes
probably indicating a melting of these textures. Therefore in this
regime the nature of the ground state is less clear. 
Restricting to vertical stripes 
we find a transition towards $d=3$ stripes both in the one-
\cite{sei04a} and three-band \cite{lor02b} model at $n_h=0.19 \sim 0.24$.
Experimentally the incommensurability $\varepsilon$ for doping
$n_h>1/8$ has been determined for LSCO \cite{yam98} and LCO
codoped with Nd \cite{tra97}. In the former material $\varepsilon
\approx 0.125$ up to $n_h=0.25$  whereas for the latter
compound the incommensurability reaches $\varepsilon \approx 0.15$
for $n_h=0.2$ which is compatible with $d=3$ stripes.   For this
solution the resonance frequency and intensity is shown with a
triangle in Fig.~\ref{fig3}.
Clearly more measurements  of the dispersion and in particular the 
resonance at doping  $n_h>1/8$ will clarify the situation.

In Fig.~\ref{fig3.5} we show the density of magnetic excitations
\begin{equation}
S(\omega)=\frac{1}{N}\sum_{q}{^{'}}  S^{eff}_q(\omega).
\end{equation}
Here the sum is restricted to the magnetic Brillouin zone for commensurate
AF order to conform to the usual experimental practice 
and $N$ denotes the number of lattice sites. 

As doping increases spectral weight is transfered from high energy to
the resonance region in good agreement with
experiment.~\cite{hay96,lor05} 
The increase of weight around the resonance energy is not a matrix
element but a density of states effect. Indeed the intensity
at the resonance decreases with doping (Fig.~\ref{fig3}) which, however, is
over-compensated by the strong Van Hove singularity produced by the 
saddle-point like dispersion around $(\pi,\pi)$. 
We have performed preliminary
calculations of $S(\omega)$ for $n_h<0.05$, 
where stripes are diagonal, which in fact indicate that at very low doping 
the density of magnetic excitations acquires again the characteristics of 
the undoped AFM.

For $n_h>0.09$, both the
roton minimum and the saddle-point ('resonance') contribute to the 
peak.  The corresponding energies are marked as vertical bars in
the panels of Fig.~\ref{fig3.5}. At $n_h=0.1$ both energies almost
coincide  and give rise to the peak in
$S(\omega)$ at $\omega = 45 $meV whereas for higher dopings they are
clearly separated.
At  $n_h=0.125$ the overall structure of $S(\omega)$ consists 
of a dominant peak around the 'resonance' frequency and several 
humps at higher energy arising from the optical bands. 
We obtain  good agreement with  the absolute intensity of the 
experiment Ref.~\onlinecite{tra04} by choosing $Z_d \,\eta_\perp \approx
0.25$ (cf. Eq. (\ref{strfac})) which is in good agreement with the estimates 
of Ref.~\onlinecite{lor05}.

\section{Conclusions}\label{conclusion}

In this paper we have studied the doping dependence of spin
excitations in high-$T_c$ materials of the LCO type based on a
stripe picture. The necessity for an understanding of these excitations  
is suggested by the direct relation found between incommensurate 
scattering parallel to the Cu-O bond and superconductivity in 
LCO compounds.\cite{wak99,wak04}

The comparison with the limited experimental data
available so far gives good agreement for both acoustic low
energy magnons and the high energy optical spin excitations. 
Also the absolute intensity of the modes is
in reasonable agreement with experiment taking into account the
results of Ref.~\onlinecite{lor05}.
The fact that we can reconcile our computations with experiments
is remarkable insofar given the simplicity of the hamiltonian and the fact
that the three model parameters $U$, $t$ and $t'$ are fixed by strong
constraints previous to the computations of spin excitations in the
doped phase.

We provide a non trivial prediction for the dependence of the velocity
of the acoustic spin-wave mode and the energy of the saddle point on doping. 
The spin-wave velocity
has a dip at $n_h=1/8$ which coincides with a similar anomaly seen in the
doping dependence of $T_c$.\cite{moo88,cra89} Although the superconducting 
anomaly is probably related to charge ordering along the stripe, 
the coincidence is intriguing.

The nature of the ground state is not so clear as one approaches  
the overdoped regime since one has to take into
account the melting of the stripes and non-Gaussian fluctuation effects
beyond RPA will become important. Also it is conceivable that
phase separation between the underdoped stripes and an overdoped Fermi
liquid occurs as mentioned in Ref.~\onlinecite{lor02b}. Since we assume a 
uniform stripe phase 
one can find evidence in favor or against this scenario by
studying the doping dependence of the saddle point energy. In the phase
separation scenario we would expect that at some point the population
of the domain walls by holes saturates thus hampering a further
increase of the saddle point energy with doping.

The GA+RPA results considerably differ from the predictions of LSWT, 
however, it is interesting that some features can be qualitatively 
reproduced by this simple theory. 
These include:
{\it i}) the difference in intensity among the two acoustic branches emerging 
from the incommensurate point,\cite{note1}
 {\it ii}) the strong intensity at the saddle point associated with the
resonance, 
 {\it iii}) the fact that the first optical 
branch in the direction perpendicular to the stripe is separated from
the acoustic branch at $(\pi,\pi)$, and  {\it iv}) the doping dependence
of the resonance energy for $n_h\le 1/8$ as shown in Fig.~\ref{fig3}.

Regarding the differences between LSWT and GA+RPA 
the most notable ones are the absence of the roton minimum in LSWT
and the different number of bands among SC and BC dispersions in LSWT vs. the
similarity of the dispersion for the two symmetries in GA+RPA.
Moreover we have found that
the dispersion of the optical branches perpendicular to the
stripes significantly differs  between GA+RPA and LSWT. For example, upon
comparing  Figs.~\ref{fig:d4} and \ref{fig:swtp2} it turns out that 
the first optical branch is inverted  with respect to its mean energy. 

Within LSWT the coupling between adjacent AFM domains is described
by an effective ferromagnetic exchange $J_F$ in case of
BC stripes (or an AFM coupling $J'$ for SC textures). 
It is interesting that the effective value of $J_F$ ($J'$), in order to
have an overall agreement with the GA+RPA results and with 
experiment, is quite small as found in other LSWT studies.~\cite{kru03}
With this small value one may wonder whether the magnetic system has 
long-range order at all or whether it is disordered due to quantum 
fluctuations.\cite{kru03} In any case we expect the system to be
close to a magnetic quantum critical point. This is in agreement with the 
experimental results of Aeppli and collaborators which indeed see 
nearly quantum critical  behavior close to the incommensurate 
wave vectors.\cite{aep97}

Regarding the RPA approach that assumes a uniform Fermi liquid as a
starting
point\cite{yin00,bri99,nor00,onu02,iik05,ito04,schny04,ere05a,erem205}  
it  has the advantage that it does not 
break $C_4$ lattice symmetry nor SU(2) spin rotational
symmetry in agreement with what one expects in overdoped
materials. 
On the other hand since the starting point is a Fermi liquid  
this  approach can not be pushed to the dilute regime or to energies of
the order of the charge transfer gap, i.e. describe charge
excitations across the Hubbard bands. 
Within this scheme one obtains a dispersion relation for spin excitations 
that differs considerably from ours. For example, there is no optical 
branch that continuously merges with the saddle point.~\cite{schny04,ere05a}
Also the low energy response  
never separates into four spots in momentum
space as  the spin-wave cones for a striped system. Instead the incommensurate spin response forms a two-dimensional structure at very low energies.
Also the integrated spectral
weight is concentrated in a much smaller energy range.\cite{nor00}
Finally for untwinned systems spectral weight is expected to appear
in both directions with some modulation due to structural 
anisotropy\cite{erem205} whereas in our theory at the lowest energies
practically all the weight is in the direction perpendicular to the
stripes.

The natural question which arises is whether the results of the present
investigation may also be relevant for an understanding of spin
excitations in YBCO. The acoustic dispersion shows a low-energy spin gap
($\Delta \sim 30$meV at optimal doping) 
and thus does not reach the incommensurate wave-vector at $\omega=0$.  
In this case it is more likely that the system is in a quantum
disordered spin phase as suggested by the ladder 
theories. \cite{voj04,uhr04,uhr05} 
One expects that the systems shows short range order with
a correlation length of the order of $\Delta/(\hbar c)\sim 5 a$
and that for energies larger than $\Delta$ the
system resembles that of an ordered phase. 
For this reason one can expect that a
computation as the present one is suitable for a description of the
universal high energy spin response. \cite{tra04,hay04,chr04} 
One encounters a similar situation in the one-dimensional Hubbard
model for large $U$ where a computation of RPA fluctuations on top 
of a spin-density wave will in many respect be in better 
agreement with the exact solution (which has no long-range order) 
than  RPA fluctuations on top of  a homogeneous Slater determinant.
Alternatively a scenario 
of fluctuating stripes where also the 
charge loses its long-range order can capture the effect of a 
spin-gap.\cite{voj05} One has to keep in mind, however, that
too much orientational fluctuations can result in an isotropic 
state which is not what is found in the 
experiment of Ref.~\onlinecite{hin04}.

In YBCO one finds  an apparent
correlation between the resonance frequency and $T_c$ in the sense
that both display a similar parabolic shape as a function of
doping so that $\omega_0\approx 5.3 kT_c$. Measurements of
$\omega_0$ have been performed from  $7\%$ underdoping up to
$2\%$ overdoping (see e.g. Refs. \onlinecite{bou02,dai01}). 
Due to the spin gap it is more  difficult to
extract the behavior of the incommensurability from the data. 
The authors in Ref.~\onlinecite{dai01}
find that $\epsilon$ already saturates for doping $n_h
\approx 0.1$ compatible with a 
stripe spacing of $d=5$ (extrapolation
to  $\omega=0$ implies a larger incommensurability similar to LCO). 
It is then reasonable to assume that the saturation corresponds to 
the situation where holes are doped into a 
stripe structure with fixed distance as in case of LCO for $n_h >1/8$. 

We find that the resonance energy as a function of doping tends to have
a saturation or a slower rate of growth with doping in the underdoped
and the overdoped regime [Fig.~\ref{fig3}a].
This is compatible with Fig.~29 in 
Ref.~\onlinecite{dai01} which reports a flattening of the
$\omega_0\,\, vs.\,\, T_c$ relation for both optimally doped {\it} and
strongly underdoped samples. In the latter case this would
correspond to the transition towards the regime where stripes are
created at distances $d>5$.

Finally our investigations suggest an explanation for the
apparent two-dimensional magnetic scattering below the resonance
in detwinned samples of YBCO.\cite{hin04}  
Arguments against a ``rigid'' stripe in this compound are based
on the assumption that in a detwinned
material stripes preferably align in the direction of the CuO
chains running along the $b^*$ direction. From spin-only stripe
models \cite{voj04,uhr04,kru03,car04} one then expects a low
energy magnetic response only in the $a^*$ direction. Instead, the
experimental data \cite{hin04,moo00nat} show an anisotropic
intensity modulation with maxima along the $a^*$ direction but
also significant weight along $b^*$. 
Lateral 
fluctuations of the stripes\cite{kiv98}  can also explain this result
which in our harmonic RPA approach are present\cite{lor03} 
but decoupled from the magnetic modes.

On the other hand our present computations show that the observed 
two-dimensional, but anisotropic intensity modulation can be explained 
by the roton like minimum in the dispersion.
In addition our spectra display an almost isotropic response for
energies well above the saddle-point in agreement with neutron scattering
results on a detwinned and underdoped YBCO sample by Stock and 
coworkers.\cite{stock05} The calculation of magnetic excitations from stripes 
within the time-dependent GA therefore yields a non-trivial crossover 
from one-dimensional behavior
at low energies to a quasi two-dimensional response at high energies
with an intermediate anisotropic but two-dimensional structure at
the roton minimum. In contrast spin-only models yield a one-dimensional
spectrum over the whole range of energies.

After this work has been submitted and posted \cite{sei05condmat} new 
INS data for an untwinned sample array of YBCO became available.~\cite{hin06} 
In our opinion the
strong anisotropy found at the lowest energies is incompatible with
uniform Fermi liquid theories.
In contrast some of our predictions based on stripes are qualitatively 
reproduced. Above the resonance
the intensity is similar for both directions in qualitative
agreement with Figs.~\ref{fig:d8d4},\ref{fig:scbc.13},\ref{fig:d4} and
in disagreement with the ladder theories.~\cite{voj04,uhr04,uhrmag}
 Moderately below the resonance there is intensity in both directions
whereas at lower energies the intensity becomes much more anisotropic
[c.f. Figs.2(a) and (b) of Ref.~\onlinecite{hin06}] which is
qualitatively consistent with a roton minimum. On the other hand our
computations do not include superconductivity, temperature effects and
damping which are clearly important in the experiments. 
In addition specific computations for bilayer systems are in process in order
to investigate the relevance of our results for YBCO materials 
more quantitatively.

\acknowledgments
G.S. acknowledges financial support from the Deutsche Forschungsgemeinschaft
and the Japanese Society for the Promotion of Science for enabling a
research visit at the IMR, Sendai where this work was completed.
We are grateful to S. Maekawa and T. Tohyama for stimulating discussions and
especially to M. Fujita, K. Yamada, J. Tranquada,  N. B. Christensen,
 D. McMorrow and B. Keimer for imparting knowledge on their experiments.

\appendix*
\section{Doping dependence of the magnetic coupling across stripes}
\label{ap1}
In this Appendix we present simple arguments to show how the 
filling of the domain walls by holes for $n_h>1/8$ affects
the effective magnetic interaction across the stripe.  
Our arguments will be based on  strong-coupling and 
we decouple the motion of charge carriers perpendicular and
along the domain walls. The kinetic energy gain of the latter
is responsible for the stability of $\nu=1/2$ half-filled stripes.
\cite{sei04a} In the following $\nu$ is fixed and we concentrate
on the transverse motion which gives rise to the magnetic
coupling between antiphase AFM domains.

{\it SC stripes:} 
For half-filled stripes  ($n_h<1/8$) the core 
will be essentially in a mixture of configurations of the kind:
\begin{eqnarray*}
  \uparrow &0         &   \downarrow  (50\%)\\
  \uparrow &\uparrow  &   \downarrow  (25\%)\\
  \uparrow &\downarrow&   \downarrow  (25\%)
\end{eqnarray*}
where we have represented three sites perpendicular to the stripe
the central one corresponding to the core. 
$0$ denotes a hole and we focused on three sites 
in which the prevailing configuration of the external spins is  
$\uparrow$ and   $\downarrow$. Since for half-filled stripes
a hole occupies a core site with probability $50\%$, the absence
of a hole implies the presence of an up or down spin 
with probability $25\%$ respectively. 
One can draw a similar picture in the
case in which the prevailing configuration of the
external spins is  $\downarrow$ and $\uparrow$.  
The hole in the core favors the AFM coupling
of the adjacent spins through virtual processes of the kind 
$0 \uparrow \downarrow 0$ where $\uparrow \downarrow$ denotes a doubly
occupied site. 
Therefore the gradual population of the core with holes for
$n_h>1/8$ should enhances the coupling 
$J'$  thus leading to an  increase of the resonance energy
as discussed in Sec. \ref{result}.

{\it BC stripes:} 
Stability of the structure requires that 
the core is composed of two sites which are coupled
by an effective ferromagnetic interaction $J_F$. The core spins 
are coupled by an effective antiferromagnetic exchange $J$ to the  
external spins.  
For the case in which the prevailing configuration of the
external spins is $\downarrow$ and $\downarrow$ (like in the upper row
of Fig.~\ref{fig:ucell}) one has for a half-filled stripe 
the typical configurations:
\begin{eqnarray*}
  \downarrow  &0\; \uparrow & \downarrow   (25\%)\\
  \downarrow  &\uparrow\; 0   & \downarrow   (25\%)\\
  \downarrow  &\uparrow\;\; \uparrow   & \downarrow   (50\%)
\end{eqnarray*}
The first two configurations favor the antiphase alignment through
processes of the kind $0\;\;\downarrow\uparrow\;\; 0\;\;\downarrow$
and  $ \downarrow \;\;0 \;\;\uparrow\downarrow\;\;0$. Since these 
process involve holes one expects an increase of the effective $J_F$
when the numbers of holes in the domain wall starts to deviate
from half-filling  for $n_h>1/8$.


\begin{thebibliography}{10}
\bibitem{notesherman}
A recent calculation within a projection-formalism for the $tJ$-model
  \cite{sherman05} is more promising in this regard.
\bibitem{brabais} We indicate the lattice associated with
the stripe solution by superlattice to distinguish is from the underling
fundamental lattice generated by ${\bf A}_1=(1,0)$ and  ${\bf A}_2
=(0,1)$.
\bibitem{noterpa} Within our scheme the RPA corresponds to the small amplitude
approximation for oscillations around the GA saddle-point in which
case the particle-hole matrix elements of $\rho_{i\sigma,j\sigma}$ may be
chosen to be independent coordinates. A full-time dependent treatment
would also imply the coupling with pairing correlations which is neglected
in the RPA (see e.g. Refs. \onlinecite{RING,BLAIZOT}).
\bibitem{resonance}
This is a simplification since we are neglecting damping effects. 
In a more accurate computation the saddle point mode will overlap with
particle-hole excitations,  making the mode much
broad.  The temperature dependent opening of a 
gap in the fluctuation spectrum of these second bosonic modes 
may produce the appearance of a sharp resonance at a smaller energy
which may be more properly called ``the resonance''. Since these effects 
are not present in our computation we do not distinguish between the
saddle point energy and the resonance energy.
\bibitem{note1}
The difference in intensity is not so obvious in the presentation of the
  results in Refs.~\onlinecite{kru03,car04} due to the different way of
  plotting the data. The weight diverges as $1/\omega$ at low frequency and
  saturates within the intensity scale of Refs.~\onlinecite{kru03,car04}. Our
  choice of plotting $ \omega \chi^{''}_q(\omega)$ avoids this problem and the
  difference in intensity becomes evident.
\end{thebibliography}
\end{document}